\documentclass[%
 aip,
 floatfix,
 amsmath,amssymb,
 reprint,%
]{revtex4-1}

\usepackage{graphicx}% Include figure files
\usepackage{dcolumn}% Align table columns on decimal point
\usepackage{bm}% bold math
\usepackage[utf8]{inputenc}
\usepackage[T1]{fontenc}
\usepackage{mathptmx}
\usepackage{etoolbox}
\usepackage{makecell}
\usepackage{xcolor}

\makeatletter
\def\@email#1#2{%
 \endgroup
 \patchcmd{\titleblock@produce}
  {\frontmatter@RRAPformat}
  {\frontmatter@RRAPformat{\produce@RRAP{*#1\href{mailto:#2}{#2}}}\frontmatter@RRAPformat}
  {}{}
}%
\makeatother
\begin{document}

\preprint{AIP/123-QED}

\title{Evaluating approaches for on-the-fly machine learning interatomic \textcolor{black}{potentials} for activated mechanisms sampling with the activation-relaxation technique nouveau}

\author{Eugène Sanscartier}
\email{eugene.sanscartier@umontreal.ca}
\affiliation{Département de physique and Regroupement québécois sur les matériaux de pointe, Université de Montréal, Case Postale 6128, Succursale Centre-ville, Montréal, Québec H3C 3J7, Canada.}
 
\author{Félix Saint-Denis}
\affiliation{Département de physique and Regroupement québécois sur les matériaux de pointe, Université de Montréal, Case Postale 6128, Succursale Centre-ville, Montréal, Québec H3C 3J7, Canada.}

\author{Karl-Étienne Bolduc}
\affiliation{Département de physique and Regroupement québécois sur les matériaux de pointe, Université de Montréal, Case Postale 6128, Succursale Centre-ville, Montréal, Québec H3C 3J7, Canada.}

\author{Normand Mousseau}
\email{normand.mousseau@umontreal.ca}
\homepage{https://normandmousseau.com}
\affiliation{Département de physique and Regroupement québécois sur les matériaux de pointe, Université de Montréal, Case Postale 6128, Succursale Centre-ville, Montréal, Québec H3C 3J7, Canada.}

\date{\today}% It is always \today, today,

\begin{abstract}
In the last few years, much efforts have gone into developing \textcolor{black}{general} machine-learning potentials able to describe interactions for a wide range of structures and phases. Yet, as attention turns to more complex materials including alloys, disordered and heterogeneous systems, the challenge of providing reliable description for all possible environments becomes ever more costly. In this work, we evaluate the benefits of using specific versus general potentials for the study of activated mechanisms in solid-state materials. More specifically, we test three machine-learning fitting approaches using the moment-tensor potential to reproduce a reference potential when exploring the energy landscape around a vacancy in Stillinger-Weber silicon crystal and silicon-germanium zincblende structure using the activation-relaxation technique nouveau (ARTn).  We find that a targeted on-the-fly approach specific and integrated to ARTn generates the highest precision on the energetic and geometry of activated barriers, while remaining cost-effective. This approach expands the type of problems that can be addressed with high-accuracy ML potentials.
\end{abstract}

\maketitle

% \begin{quotation}
% The ``lead paragraph'' is encapsulated with the \LaTeX\ 
% \verb+quotation+ environment and is formatted as a single paragraph before the first section heading. 
% (The \verb+quotation+ environment reverts to its usual meaning after the first sectioning command.) 
% Note that numbered references are allowed in the lead paragraph.
% %
% The lead paragraph will only be found in an article being prepared for the journal \textit{Chaos}.
% \end{quotation}

\section{\label{intro}Introduction}

As computational materials scientists turn to attention to ever more complex systems, they are faced with two major challenges : (i) how to describe correctly their physics and (ii) how to reach the appropriate size and time scale to capture the properties of interest. The first challenge is generally solved by turning to \textit{ab initio} methods,~\cite{PhysRev.140.A1133} that allow the solution \textcolor{black}{Schrödinger's} equation with reasonably controlled approximations. Theses approaches, however, suffer from $N^4$ scaling which limits their application to small system sizes and short time scales. The second challenge is met by a variety of methods that cover different scales. Molecular dynamics~\cite{lindahl2008molecular}, for example, which directly solves Newton's equation, accesses typical time scales between picoseconds and microseconds, at the very best. Other approaches, such as lattice~\cite{voter1984transition, voter2007introduction} and off-lattice kinetic Monte-Carlo~\cite{henkelman2001,el-mellouhi_kinetic_2008}, by focusing on physically relevant mechanisms, can extend this time scale to seconds and more, as long the diffusion takes place through activated processes. Even though these methods are efficient, each trajectory can require hundreds of thousands to millions of forces evaluations, becoming too costly with \textit{ab initio} approaches, forcing modellers to use empirical potentials in spite of their incapacity at describing correctly complex environments.

Building on \textit{ab initio} energy and forces, machine-learned potentials (MLP)~\cite{behler_generalized_2007,bartok2013representing, thompson2015spectral, shapeev2016moment} open the door to lifting some of this difficulties, by offering much more reliable physics as a small fraction of the cost of \textit{ab initio} evaluations. 

Since their introduction, ML potentials have been largely coupled with MD, focusing on the search for universal potentials able to describe a full range of structures and phases for a given material~\cite{sivaraman2021automated, kang2020large, sivaraman2020machine}. As we turn to more complex systems such as alloys and disordered and heterogeneous systems, it becomes more and more difficult to generate such universal potentials, since the number of possible environments grows rapidly with this complexity. In this context, the development of specific potentials, with on-the-fly learning that makes it possible to adapt to new environments, becomes a strategy worth exploring. 

In this work, we focus on the construction of machine-learned potentials adapted to the sampling of energy landscape dominated by activated mechanisms, i.e., solid-state systems with local activated diffusion and evolution. \textcolor{black}{This kinetics is associated with aging and relaxation of disordered materials~\cite{beland_replenish_2013}, sluggish diffusion in concentrated alloys~\cite{osetsky2018existence} and defect diffusion in complex environments~\cite{restrepo2018carbon}. A correct computational sampling,  using open-ended methods such as the activation-relaxation technique (ART)~\cite{barkema_event-based_1996} and its revised version (ART nouveau or ARTn)~\cite{malek_dynamics_2000,jay_activationrelaxation_2022}, requires a precise description of local minima and of the landscape surrounding the first-order saddle points that characterize diffusion according to the transition-state theory (TST)~\cite{truhlar1996current}. These barriers can be high --- reaching many electron-volts --- and involve strained configurations that can be visited only very rarely with standard molecular dynamics.}
\textcolor{black}{Yet, because of this need for long-time evolution, these problems have been largely out of reach of \textit{ab initio} approaches and have been studied, until now, mostly with empirical potentials.}

\textcolor{black}{More specifically, we compare three machine learning approaches to train a Moment Tensor Potential (MTP)~\cite{shapeev2016moment, novikov2020mlip} for the diffusion of a vacancy in silicon and silicon-germanium alloy as sampled with ARTn. The first approach consists in a pure MD learning, fitted at various temperatures, following steps that echo the work of Novoselov \textit{et al.}~\cite{novoselov2019moment}; the second approach adds an on-the-fly training during ARTn runs and the third one focuses on a purely on-the-fly training during ARTn runs. While on-the-fly learning is not a new approach and have been use before in various context, they have never been compared for the specific application of activated kinetics in solid state systems.}

\textcolor{black}{To generate the statistics necessary to offer solid conclusions, we select to use the Stilliger-Weber empirical potential both for Si~\cite{stillinger1985computer} and SiGe~\cite{ethier1992epitaxial}. Clearly, the machine-learned potentials generated here have therefore limited physical relevance, these are sufficient to allow us to assess the relative quality of our three approaches.}

Results underline the efficiency gain in developing targeted ML potentials for specific applications, comparing the cost of fitting Si with SiGe, it also shows the rapid increase in computation complexity associated with moving from element to alloy systems, which emphasizes the usefulness of a specific approach such as the one applied here to activated processes. 

\section{Methodology}

\subsection{ML Potential}

The Moment Tensor Potential (MTP)~\cite{shapeev2016moment,novikov2020mlip} is a linear model of functions $B_\alpha(\mathfrak{r}_i)$ built from contractions of moment tensor descriptors defined by the local neighborhood relative position $\mathfrak{r}_i$ of atom $i$ within a sphere of influence of radius $r_c$ respecting a set invariances. This model has been shown to be fast while giving accuracy on the order of $\sim$meV/atom and requiring few hundreds to thousands of reference potential calls~\cite{zuo2020performance} on-the-fly.

MTP have been used on a wide variety of problems including on-the-fly MD simulation~\cite{novoselov2019moment, novikov2020mlip, podryabinkin2017active}, search and minimization of new alloys~\cite{podryabinkin2019accelerating, gubaev2019accelerating} and diffusion processes~\cite{novoselov2019moment} on systems counting one or multiple species. In the following we offer a summary of the method; more information on MTP is available in Ref.~\onlinecite{novikov2020mlip}.

MTP approximates atomic configuration energy as sum of local contributions. A local contribution is obtained through a sum over the included basis $\{B_\alpha(\mathfrak{r}_i)\}$ as a linear combination of $B(\mathfrak{r}_i)$ and $\xi_\alpha$,
\begin{equation}\label{eq:Vcontrn}
    V(\mathfrak{r}_i)=\sum_{\alpha=1}^{m} \xi_\alpha B_\alpha(\mathfrak{r}_i).
\end{equation}

The ``level'' of a potential gives the number of different possible tensor $M_{\mu, \nu}\left(\mathfrak{r}_{i}\right)$ descriptors. The $B_\alpha(\mathfrak{r}_i)$ functions of Eq.~\ref{eq:Vcontrn} are constructed by a tensorial contraction of different $M_{\mu, \nu}\left(\mathfrak{r}_{i}\right)$ and the number of different tensorial contraction sets $m$ in Eq.~\ref{eq:Vcontrn}. More information on MTP is available in Ref.~\onlinecite{novikov2020mlip}.

The total energy of a N-atom configuration ($\mathfrak{R}$) is then given by the sum of N local contributions,
\begin{equation}\label{totenen}
    E(\mathfrak{R})=\sum_{i=1}^{N} V(\mathfrak{r}_i) = \sum_{i=1}^{N} \sum_{\alpha=1}^{m} \xi_\alpha B_\alpha(\mathfrak{r}_i),
\end{equation}
and the forces are obtained by taking the gradient of this quantity,
\begin{equation}
    \textbf{F}(\mathfrak{R})=-\nabla \sum_{i=1}^{N} \sum_{\alpha=1}^{m} \xi_\alpha B_\alpha(\mathfrak{r}_i).
\end{equation}

The parameters $\xi_\alpha$ are obtained by minimizing the loss function:
\begin{equation}\label{eq:loss}
    \sum_{\mathfrak{R} \in \mathrm{A}} \left[ w_e \left(E(\mathfrak{R})  - \hat{E}(\mathfrak{R})\right)^2 + w_f \sum_i^N \left|\textbf{f}_i(\mathfrak{R}) - \hat{\textbf{f}}_i(\mathfrak{R})\right|^2 \right] \rightarrow \min_\xi.
\end{equation}

Here $\mathrm{A}$ is the training set made of configurations with known energy and forces. The goal is to minimize the difference between $E(\mathfrak{R})$, $\textbf{f}_i(\mathfrak{R})$(real value) and $\hat{E}(\mathfrak{R})$, $\hat{\textbf{f}}_i(\mathfrak{R})$(predicted by model), respectively, for all element in $\mathrm{A}$. Weights on contribution from energy and forces ($w_e$ and $w_f$) are set to one.

\subsection{Learning On-The-Fly Tools}
\color{black}
On-the-fly atomic machine learning potential (OTF) involves the repeated training of the model potential as new atomic environments are generated through various procedures. 

Following the work of Shapeev and collaborators~\cite{novikov2020mlip}, the reliability of the potential to evaluate a given configuration is done using the D-optimality criterion which state that the best training set is the one with maximal volume.
The volume can be seen as a measure of the domain of the training set where the model allows reliable predictions. Assuming that the training set is of maximal volume, any configuration within or beyond this volume is inside or outside the known domain of the model, respectively, which we interpret as interpolation and extrapolation.
We grade with the D-optimality criterion of a given configuration by assessing the model reliability. For this, Shapaeev \textit{et al.}  introduce a selection algorithm (MaxVol) that tests whether or not this configuration should be added to the training set or replace a configuration already in it
While a detailed description can be found in Ref.~\cite{podryabinkin2017active}, we provide here a brief summary of the retained approach.

The selection and extrapolation-grade algorithm can be applied using either a local-energy or a global-energy descriptor.

The local-energy descriptor is presented as a rectangular matrix $\mathrm{G}_{N\times\mathrm{m}}$ formed by the basis elements $B_\alpha(\mathfrak{r}_i)$ associated with the neighborhood $\mathfrak{r}_i$ of all $N$ atoms:
\begin{equation}\label{eq:G}
    \mathrm{G}=
    \left(
    \begin{array}{ccc}
    B_{1}(\mathfrak{r}_{1}) & \ldots & B_{m}(\mathfrak{r}_{1}) \\
                     \vdots & \ddots & \vdots \\
    B_{1}(\mathfrak{r}_{N}) & \ldots & B_{m}(\mathfrak{r}_{N})
    \end{array}
    \right).
\end{equation}

For a given configuration, the global-energy description reduces this information to a vector $\mathbf{g}$,
\begin{equation}
    \mathbf{g} = \left( \begin{array}{ccc} b_{1}(\mathfrak{R}) & \ldots & b_{m}(\mathfrak{R}) \end{array} \right),
\end{equation}
where each term, $\{b_\alpha(\mathfrak{R})\}$ is a sum over all neighborhoods for a specific basis element $\{B_\alpha(\mathfrak{r}_i)\}$:
\begin{equation*}
    \{b_\alpha(\mathfrak{R})\} = \sum_{i=0}^{N} \{B_\alpha(\mathfrak{r}_i)\}.
\end{equation*}

For the global-energy descriptor of a given configuration with a training set $\mathrm{A}$, solving for $c_{ij}$, in
\begin{equation}\label{eq:max_solve}
    \left( \begin{array}{ccc} c_{11} & \ldots & c_{1m} \end{array} \right) \mathrm{A} = \mathbf{g}.
\end{equation}
Assuming that $\mathrm{A}$ is of maximal volume,  $0 \leq c_{ij} \leq 1$. Conversely, if one find $c_{ij} > 1$ then $\mathrm{A}$ is not of maximal volume. The later case mean that the model is extrapolating on this configuration and that updating $\mathrm{A}$ with this configuration will increase its volume.
The extrapolation grade, $\gamma$, is then defined as the largest component of $c_{ij}$,
\begin{equation}\label{eq:grade}
    \gamma(\mathfrak{R})=\max \left|c_{ij}\right|.
\end{equation}

The same approach is used for the local-energy description, applying Eq.~\ref{eq:max_solve} with the rows of matrix $\mathrm{G}$ rather than the vector $\mathbf{g}$. For non-linear and other ML potentials, in Eq.~\ref{eq:G}, one adapt this algorithm performing the ML model’s gradient with respect to its parameters.

In practice, for $\gamma(\mathfrak{R})$ below a certain threshold $\gamma_0$, the model interpolation is reliable,  while for $\gamma_0 < \gamma(\mathfrak{R}) < \gamma_{max}$, the model cannot be applied with confidence, but can be adapted by adding this configuration to the training set. When  $\gamma(\mathfrak{R}) > \gamma_{max}$, the configuration is too far from the training set and it is rejected as the model cannot be adapted with confidence. In this work, we set $\gamma_0 = 1.1$ and $\gamma_{max} = 2.2$, taking the lower values of the parameters studied in Ref.~\cite{podryabinkin2017active}, unless specified otherwise.
\color{black}

\subsection{On-The-Fly Learning Cycle Workflow}\label{workflow}

Our workflow is similar to that of Ref.~\onlinecite{novikov2020mlip}, with main differences  discussed in Section~\ref{sub:details}. We follow the same general machine-learning on-the-fly workflow for all sampling approaches tested here.

\begin{figure}[h]
\centering
\includegraphics[width=\linewidth]{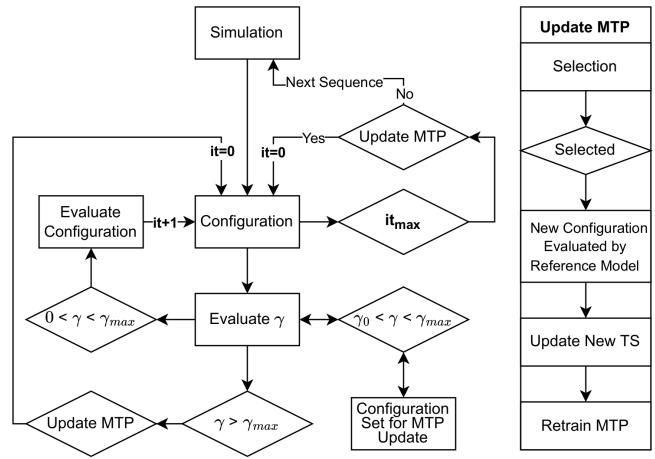}
\caption{\label{fig:otf} On-the-fly machine learning workflow used with MD and ARTn (on the left). A potential update can take place at two points: when the sequence ends or when $\gamma > \gamma_{max}$. The updating procedures are given in the box on the right.}
\end{figure}

We split each simulation in one or multiple sequences of atomic configurations generated using either MD or ARTn. Each run \textcolor{black}{unfolds} as follows (see Fig.~\ref{fig:otf}):
\begin{enumerate}
    \item Launch a sequence during which configurations are generated according to a sampling algorithm (MD or ARTn).
    
    At each iteration step the extrapolation-grade $\gamma$ is evaluated.
        \begin{enumerate}
            \item If $0 < \gamma < \gamma_{max}$, the energy and forces of the configuration are evaluated with MTP; 
            
            \item if $\gamma_{0} < \gamma < \gamma_{max}$, the configuration is set aside for an update of MTP parameters;
            
            \item else if $\gamma > \gamma_{max}$, energy and forces of the configuration are not evaluated with MTP and the configuration is not kept for update. The sequence is stopped and we go directly to the update step (step 3).
        \end{enumerate}

    \item Move on next to the iteration in the sequence (step 1).
    
     \item The model is updated, if at at least one configuration as been set aside for an update of MTP (i) at the end of a sequence or (ii) at any moment during the sequence if $\gamma > \gamma_{max}$.
     
     \item If there is an update, restart a new sequence (go to step 1), else stop if no configuration with $\gamma > \gamma_{0}$ has been set aside during the predefined maximum length of the sequence.
\end{enumerate}

The moment tensor potential model update is defined as follows (see Fig.~\ref{fig:otf}, right-hand side):
\begin{enumerate}
    \item A selection is made from the set aside configurations (with $\gamma > \gamma_{0}$) using MaxVol~\cite{podryabinkin2017active}.
    \item Each selected configuration is evaluated by the reference model
    \item The training set is updated with the new evaluated configurations
    \item The moment tensor potential is fitted on the new training set accordingly to Eq.~\ref{eq:loss}
\end{enumerate}
More details of this procedure can be found in Ref.~\onlinecite{podryabinkin2017active}.

\subsection{MD and ARTn}

Two sampling approaches are used to generate a sequence of configurations: (1) molecular dynamics (MD) as implemented within LAMMPS~\cite{LAMMPS} and (2) the activation-relaxation technique nouveau (ARTn) algorithm developed by Mousseau and collaborators~\cite{barkema_event-based_1996,malek_dynamics_2000,jay2022activation}. Since MD is well known, we only give below a brief summary of ARTn.

ARTn is designed to explore the potential energy landscape of atomic systems through the identification of local transition states connecting nearby local minima. Its workflow can be summarized in three main steps (see, for a recent in depth discussion of the ARTn version used in this work, see Ref.~\onlinecite{jay2022activation}):

\begin{enumerate}
    \item \textit{Leaving the harmonic well}: starting from an energy minimum, an atom and its neighbours are moved iteratively in a direction selected at random until a direction of negative curvature on the potential energy surfaces, $\mathbf{d}(\lambda_\mathrm{min})$ with $\lambda_\mathrm{min}$, the lowest eigenvalue of the Hessian matrix, smaller than zero, emerges; this indicates the presence of a nearby first-order saddle point;
  
    \item \textit{Converging to a first-order saddle point}: the system is then pushed in the direction of negative curvature $\mathbf{d}(\lambda_\mathrm{min})$ while the force is minimized in the perpendicular plane, until the total force $F$ passes below a threshold near $F_0$, which indicates the saddle point have been reached;

    \item \textit{Relaxing into a new minimum}: the system is then pushed over the saddle point and relaxed into a connected new minimum.
\end{enumerate}

At each step $\lambda_\mathrm{min}$ and $\mathbf{d}(\lambda_\mathrm{min})$ are found using an iterative Lanczos method~\cite{lanczos1950iteration,malek2000dynamics,jay_activationrelaxation_2022}. Perpendicular relaxation during activation and global minimization are done using the Fast Inertial Relaxation Engine (FIRE) algorithm~\cite{bitzek2006structural}.

Generated events are accepted or rejected according to the Metropolis algorithm, where the acceptation probability $p$ is given by
\begin{equation}\label{eq:metropolis}
    p = \min \left( 1, e^{-\beta \Delta E} \right),
\end{equation}
with $\Delta E = E_{\mathrm{saddle}} - E_{\mathrm{minimum}}$, the energy difference between the saddle and a connected minima and $\beta = 1/k_B T$ where $k_B$ is the Boltzmann factor and $T$ is a fictitious temperature, since thermal deformations are not taken into account. Potential energy landscape exploration consist of generating a number of event.

\subsection{Systems studied}

The fitting approaches are tested on two physical systems: (i) a Si diamond structure with Stillinger-Weber as a reference potential~\cite{stillinger1985computer}; and (ii) a SiGe zincblende structure using the Stillinger-Weber potential with parameters from Ref.~\cite{ethier1992epitaxial}. Both models count 215 atoms and a vacancy. These potentials as selected as they are computationally light and facilitate the accumulation of statistics at the level needed here; \textcolor{black}{to obtain physically-relevant results, one should train on DFT and not on empirical potentials.}

The Si system is fitted with a ML potential set at level 16, with 92 moment tensor functions ($B(\mathfrak{R})$, Eq. \ref{eq:Vcontrn}). For SiGe, a potential at this level (16) generates errors on the barrier of the order of 0.5~eV, which indicates that a richer set of parameters is needed to describe the chemical diversity and a level 20 is chosen for this system, with 288 moment tensor functions. The relation between the number of moment tensor functions for Si and energy error is presented in Supplemental Fig.~1.

\subsection{Fitting approaches}\label{sub:details}

To evaluate the reliability of the various on-the-fly approaches to reproduce the reference potential on configurations of interest for complex materials, the training set is limited to structures visited during MD or ARTn simulations within the conditions described below. No additional information regarding alternative crystalline structures, defects, surfaces, pressure, etc. is provided. 

For each of these two systems, we compare the following approaches:

\begin{enumerate}
    \item{ML-MD}: The MTP potential is train OTF on MD simulations. The potential is then evaluated, \textit{without further update}, in ARTn simulation.
    
    \item{OTF-MDART}: Starting from the ML-MD generated potential, the MTP is re-trained following the OTF procedure during ARTn simulations.

    \item{OTF-ART}: Training of the potential is done uniquely during ARTn runs with OTF.
\end{enumerate}

The ML-MD approach is in line with Ref.~\onlinecite{novoselov2019moment} where a potential is trained OTF during MD. However, while the potential is trained with MD, its accuracy is  evaluated during ARTn activated process search.

\subsubsection{ML-MD: simulations details}

Nine sets of MTP ML-MD potentials are developed and trained independently during NVT MD simulations. Each set is trained at one specific simulation temperature ranging from 300~K to 2700~K by step of 300~K and starting from the same 215~atom crystalline structure with a vacancy. Each set consists of ten independently constructed MTP potentials for statistical purpose.

Training takes place on a series of sequences, each run for a maximum of 100~ps, with steps of 1~fs, with an average of 75~ps per cycle. MTP potentials require about $34\pm14$ and $93\pm43$ learning cycles for Si and SiGe to be converged: the MTP potential is considered having learned the potential when no configuration generated during a 100~ps second is found in the extrapolating zone of the potential (with $\gamma > \gamma_{0}$).

As long as this is not the case, the sequence is restarted from the same initial structure with different initial velocities. To facilitate convergence, ML-MD potentials are fitted  over three sets of progressively more restricted reliability extrapolation parameter $\gamma_0$. Moreover because MD leads to global deformation, the extrapolation is computed using global descriptors (see tab. \ref{tab:hyper-param}).

The final potential is then evaluated, in a fixed form, in ARTn simulations.

\begin{table}[h]
\caption{Extrapolation and selection hyper-parameter values used for the three on-the-fly approaches used in this work.}\label{tab:hyper-param}
\begin{ruledtabular}
\begin{tabular}{l c c c r}
approach:          &  & $\gamma_{0}$ & $\gamma_{max}$ & \makecell{grade-\\mode}  \\ \hline \\
ML-MD              &  & 5.5/3.3/1.1  & 60/10/2.2      & global   \\
OTF-MDART          &  & 1.1          & 2.2            & local    \\
OTF-ART            &  & 1.1          & 2.2            & local    \\
\end{tabular}
\end{ruledtabular}
\end{table}

\subsubsection{OTF ARTn simulations details}

Each ARTn simulation is launched for 1500 events, with 24 parallel independent searches, for a total of 36~000 generated events. For ARTn, a sequence is either a search for a saddle point (successful or failed) or a minimization from the saddle to minimum. 

At each point, 24 sequences are generated in parallel, and the configuration selected for an update of the potential is made on the combined set of configurations to generate one training set. Sequence are restarted from the last accepted position or, in the case of the vacancy in Si, the ground state. When an activation step generates a configuration with $\gamma(\mathfrak{R}) > \gamma_{max}$, it is relaunched with the same initial deformation. As with MD, ten independent ARTn runs are launched for statistics.

In the bulk, diffusion of the vacancy in Si takes place through a symmetric mechanism bringing the vacancy from one state to an identical one so all ARTn event searches are effectively started from the same state. Starting from a zincblende structure, SiGe evolves according to an accept-reject Metropolis 
with a fictitious temperature of $0.5$~eV~\cite{mousseau_exploring_1999}. Since the configurations explored by ARTn are locally deformed; the extrapolation grade for ARTn generated configurations used for the OTF-MDART and OTF-ART approaches are evaluated with the local descriptors.

\subsection{Analysis}

Following the standard approach, the error is computed on the energy and force differences between the MLP and reference potentials computed on the same structures. Here, however, this error is only measured on configurations generated during the ARTn procedure.

For the energy:
\begin{equation}
    \Delta E = | E_{MLP}(X_{MLP}) - E_{ref}(X_{MLP})  |,
\end{equation}
and, for the forces:
\begin{equation}
    \Delta F =  \frac{1}{N} \sum_{i=0}^{N} \sqrt{ \| \mathbf{f}_{MLP}^{(i)}(X_{MLP}) - \mathbf{f}_{ref}^{(i)}(X_{ref}) \|^{2}},
\end{equation}
where the positions $X_{MLP}$ are obtained from a simulation run with the machine-learned potential and the energy on this exact configuration is computed with the reference and the machine-learned potentials. The same is done for the error on forces.

Since this work is focused on the correct description of first-order transition states, we also compute the minimum and saddle barrier positions and energy convergence errors ($\Delta X_{\text{conv}}$, $\Delta E_{\text{conv}}$) as 
\begin{eqnarray}\label{eq:conv}
    \Delta X_{\text{conv}} &= \sqrt{\sum_{i=0}^{N} \| \mathbf{x}^{(i)}_{MLP} - \mathbf{x}^{(i)}_{ref} \|^{2}}, \\
    \Delta E_{\text{conv}} &= | E_{MLP}(X_{MLP}) - E_{ref}(X_{ref})  |,
\end{eqnarray}
where $X_{MLP}$ and $X_{ref}$ are the positions corresponding to minimum or saddle point as defined by the MLP and the reference potentials respectively, with $E_{MLP}(X_{MLP})$ and $E_{ref}(X_{ref})$ the corresponding energies; by definition, forces are zero at these points defined by the respective potentials.

While $X_{MLP}$ and $E_{MLP}(X_{MLP})$ are obtained on the ARTn trajectories, $X_{ref}$ and  $E_{ref}(X_{ref})$ are obtained after reconverging the minima or the saddle point using the reference potential starting from $X_{MLP}$ and following the ARTn procedure.

From an energy barrier $\delta E(X)$, the energy barrier error $\Delta \delta E_{barrier}$ is given by
\begin{equation}
\label{eq:barrier}
    \Delta \delta E_{barrier} = | \delta E_{MLP}(X_{MLP}) - \delta E_{ref}(X_{ref})  |.
\end{equation}

If no trend is observed between the different temperatures where potentials are trained, we calculate their average and deviation in order to to effectively compare them with other approach.

\section{Results}

\begin{figure}[h]
\centering
\includegraphics[width=\linewidth]{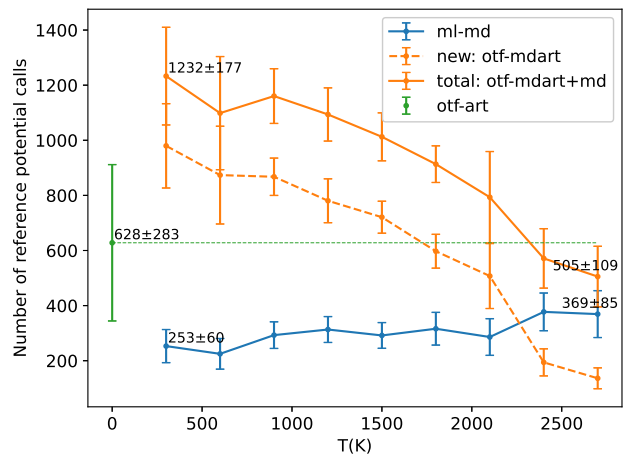}
\caption{\label{fig:si-counts} Number of calls to the reference potential for each of the machine-learned potentials developed for Si as function of the temperature referring to the one used during MD training. Since configurations are relaxed to zero K in ARTn simulations, there is no associated temperature for this procedure. Vertical bars represent the standard deviation computed on ten independent realisations.}
\end{figure}

\begin{figure}[h]
\centering
\includegraphics[width=\linewidth]{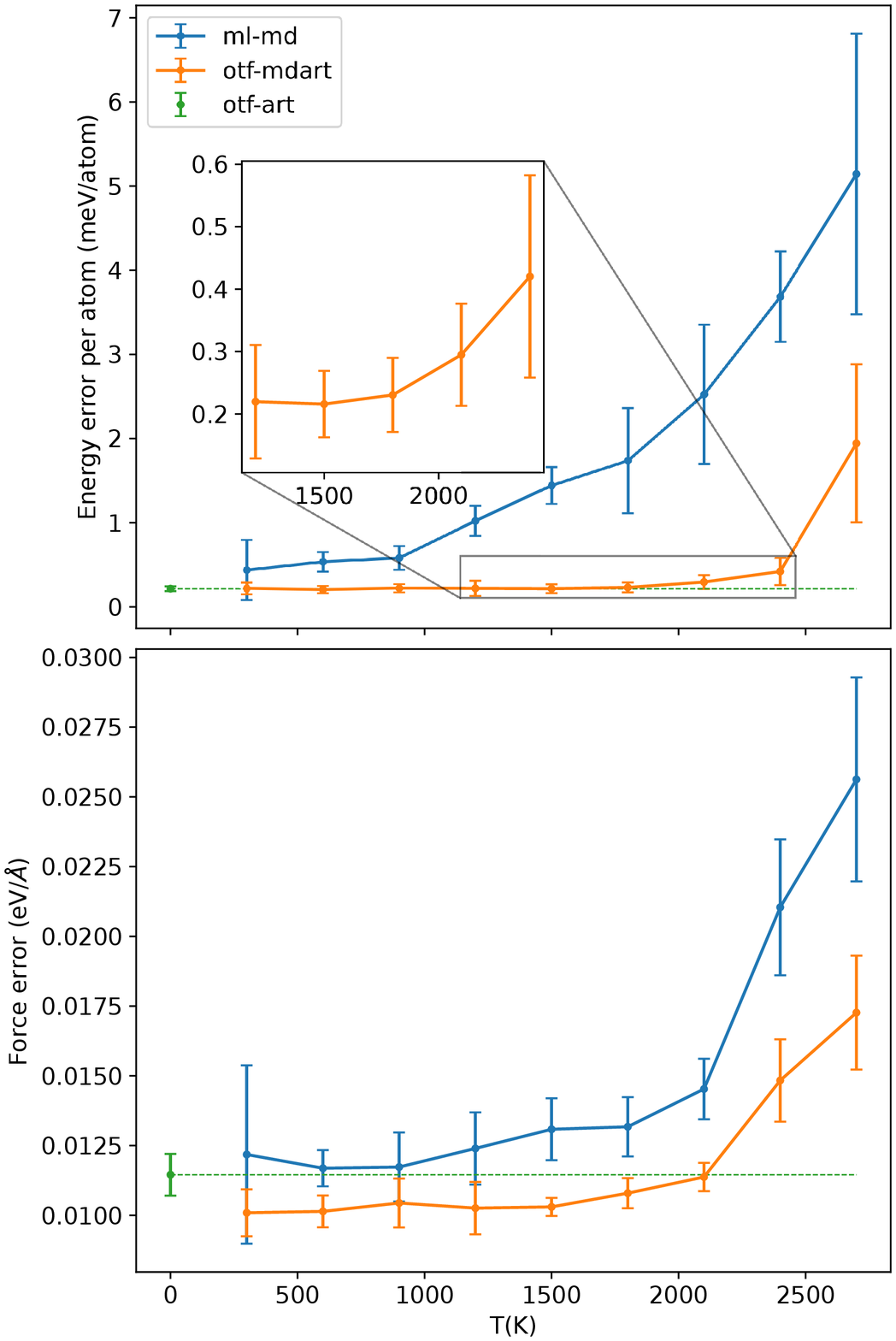}
\caption{\label{fig:si-error_mean_valid} Average energy (top) and mean absolute forces (bottom) errors per atom for Si measured over all configurations generated along pathways in ARTn for the three approaches. Temperature refers to the one used during MD training. Vertical bars represent the standard deviation computed on ten independent realisations.}
\end{figure}

\begin{figure}[h]
\centering
\includegraphics[width=\linewidth]{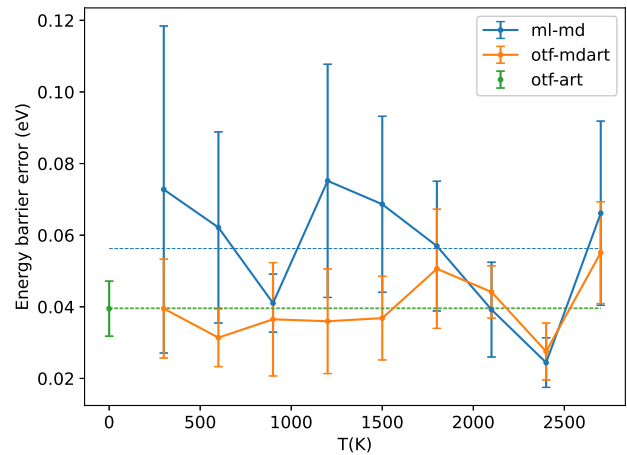}
\caption{\label{fig:si-barrier_error-all} Average energy barrier error for Si as defined by Eq.~\ref{eq:barrier} for all events generated in ARTn for the three approaches. Temperature refers to the one used during MD training. Vertical bars represent the standard deviation computed on ten independent realisations.}
\end{figure}

\begin{figure}[h]
\centering
\includegraphics[width=\linewidth]{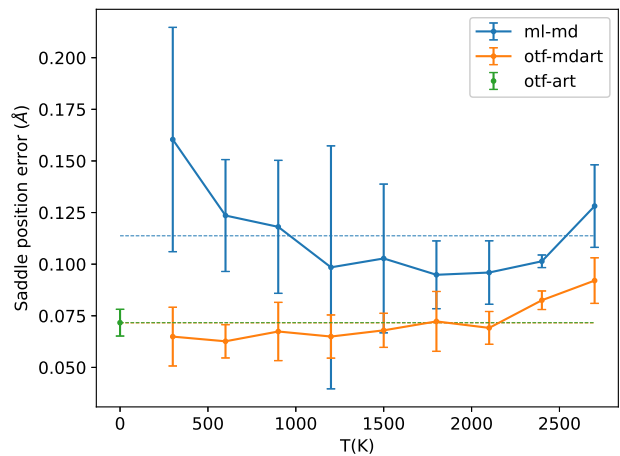}
\caption{\label{fig:si-pos_saddle} Mean position error on all saddle point for Si. Temperature refers to the one used during MD training. Vertical bars represent the standard deviation computed on ten independent realisations.}
\end{figure}

In this section,  we first examine results for a vacancy in  c-Si to establish the methods then consider the same approaches on the more complex SiGe alloy.

\subsection{ML-MD}

The ML-MD approach serves as a benchmark to assess the efficiency of the various approaches in sampling energy barriers and diffusion mechanisms. Here, ten independent ML potentials are generated through on-the-fly MD simulations at 9 different target temperatures ranging from 300 to 2700~K by step of 300~K and require between $253\pm60$, at 300~K, and $369\pm85$ evaluations of the  reference potential, at 2700~K, to complete learning cycles~(see Fig.~\ref{fig:si-counts}).

For the purpose of this work, the quality of the ML-MD potential is evaluated on configurations generated with ARTn as local activated events associated with vacancy in a crystalline environment are generated. To avoid non-physical results, when a ARTn-generated configuration shows a $\gamma > 200$, the configuration is rejected, the event search is stopped and a new event search is launched from the same initial minimum. 

Fig.~\ref{fig:si-error_mean_valid} shows the standard validation error on energy and forces calculated over \textit{all} configurations generated along pathways for the 36~000 successful events and 10~080 failed saddle searches (a success rate of 78~\%). The error on energy increases almost exponentially with the sampling temperature, ranging from $0.44\pm0.36$~meV/atom at 300~K to $5.1\pm1.7$~meV/atom at 2700K. The error on forces is essentially constant at 0.0123~eV/\r{A}, on average, between 300 and 1800~K, and increases rapidly at high temperature, to reach 0.0256~eV/\r{A} at 2700~K.

Since, the focus of this work is on transition states, Fig.~\ref{fig:si-barrier_error-all} displays the error on the energy barriers as a function of MD-fitting temperature, computed with Eq.~\ref{eq:conv} and averaged over all generated barriers. This error is relatively uncorrelated of the MD temperature simulation with an average of $0.056 \pm 0.022$~eV, with minimum error of $0.024 \pm 0.01$~eV at 2400~K and maximum of $0.08 \pm 0.03$~eV at 1200~K.  This error is lower than that for a general point on the energy landscape (Fig.~\ref{fig:si-error_mean_valid}) in part because it is computed as a difference between saddle and initial minimum. 

Errors on the position of the saddle point, associated with the capacity to reproduce correctly their geometry, are given in Fig.~\ref{fig:si-pos_saddle}. The top panel indicates the average distance between saddle points converged with the reference and the ML potentials: it decreases from $0.16\pm0.05$~\r{A} at 300~K to a minimum of $0.09\pm0.02$~\r{A} between 1500 and 2100~K, going up at the two highest temperatures (2400 and 2700~K).

Overall, this straightforward fitting approach based on constant-temperature MD runs provides accurate diffusion barriers, ranging from 0.51 to more than 4~eV, for a vacancy in crystalline silicon at a low computational costs (263 to 369 evaluations of the reference potential).  

\subsection{Revisiting ML-MD potential in ARTn: the OTF-MDART adjusting approach}

To evaluate the possibility of improving on ML-MD potentials for activated events, potentials are on-the-fly re-trained during ARTn learning cycles (OTF-MDART). Fig.~\ref{fig:si-counts} gives the number of calls to the reference potential for this procedure during the ARTn runs (dashed orange line) as well as the total number of calls, including those made during ML-MD fitting  (solid orange line). The number of calls during ARTn learning cycles  ranges from $979\pm153$ at 300~K to to $136\pm38$ at 2700~K for a total of $1232\pm177$ to $505\pm109$ respectively, when  including ML-MD calls.

The error on energy and forces remains correlated with the ML-MD temperature: it is higher when the error is higher at ML-MD trained temperature. This correlation is particularly strong when retraining MD potentials fitted between 1500 and 2700~K (Fig.~\ref{fig:si-error_mean_valid}, solid orange line). Error on energy for OTF-MDART is almost constant between 300 and 2400 K, at 0.22~meV/atom, rising to 1.9~meV/atom at 2700~K, lower by 50 to 63~\% than ML-MD.  As similar improvement is observed on the forces, which range from 0.0103~eV/\r{A}, on average, between 300 and 1800~K,  increasing to  0.0173~eV/\r{A} at 2700~K, representing a 16~\% to 32~\% decrease in error.

\begin{table}[h]
\caption{\label{tab:si-barr-pos-error-per} Average energy barrier error and mean position error on all saddle point for Si. The average error for ML-MD and OTF-MDART training is taken over all temperature sets. The standard deviation computed on all temperature sets.}
\begin{ruledtabular}
\begin{tabular}{l c c c r}
\multicolumn{1}{l}{Errors} & \multicolumn{1}{c}{} & ML-MD           & OTF-MDART       & OTF-ART         \\ \hline \\
$\Delta \delta E_{barrier}$ (eV)             &    & 0.056$\pm$0.022 & 0.040$\pm$0.012 & 0.039$\pm$0.008 \\
$\Delta X_{\text{conv}}$ (\r{A})             &    & 0.114$\pm$0.029 & 0.072$\pm$0.010 & 0.072$\pm$0.006 \\
\end{tabular}
\end{ruledtabular}
\end{table}

Between 300 and 1500~K, retrained potentials with OTF-MDART show more constant energy barrier errors than pure ML-MD models (Fig.~\ref{fig:si-barrier_error-all}), with an error  of about $0.036$~eV (OTF-MDART) vs average of $0.064$~eV (ML-MD) a 44~\% improvement. At the highest temperature --- 1800 to 2700~K, however, as OTF-MDART calls for less learning cycles, errors and fluctuations are not reduced with respect to ML-MD. Interestingly, though, improvements on the saddle position is observed at all temperatures for OTF-MDART (Fig.~\ref{fig:si-pos_saddle}) with an average error of $0.072\pm0.010$~\r{A}.

Overall, by retraining ML-MD potential in ARTn, errors are reduced and results are more consistent, i.e., error distributions are narrower, irrespective of the temperature used in the initial MD training. This additional retraining leads to a 50~\% to 96~\% decrease in energy error~(Fig.~\ref{fig:si-error_mean_valid}), a 29~\% improvement for average energy barrier errors~(Tab.~\ref{tab:si-barr-pos-error-per}) and a 37~\% reduction on mean saddle positions errors but with an additional number of calls to the reference potential increasing between 37 to 490~\%. 

\begin{figure}[h]
\centering
\includegraphics[width=\linewidth]{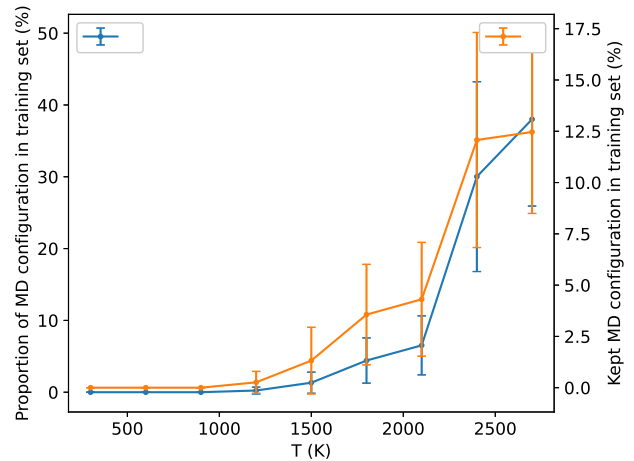}
\caption{\label{fig:origin} Fraction of original MD configurations (left scale) and total number of MD configurations (right scale) remaining in the final training set (TS) for Si. Temperature refers to the one used during MD training. Vertical bars represent the standard deviation computed on ten independent realisations.}
\end{figure}

These results can be understood by looking at the fraction of MD-generated configurations that remain in the training set at the end of the simulation (Fig.~\ref{fig:origin}): 
at temperatures between 300 and 1200~K, none of the ML-MD configurations remain in the final training set; this proportions goes from from 1.3 to 38~\% between 1500 and 2700~K (left-hand axis, blue line). At these temperatures, the system melts and generates a wider range of configurations. Since these configurations are far from ARTn-generated configurations, the selection algorithm keeps them in the set even though they do not help reduce errors for the configurational space of interest with ARTn.

\subsection{The OTF-ART adjusting approach}

Given the results for OTF-MDART, we now turn to an OTF approach entirely integrated in ARTn, in an attempt to increase accuracy, and reduce the cost and waste of evaluations of the reference potential.

Ten independents on-the-fly ML potential are generated entirely in ARTn for a total of 36~000 events starting from the same initial minimum. Each potential is trained initially from the same one configuration (the initial minimum), in the training set. Each parallel event search goes trow a learning cycle if needed and as the simulation progresses learning cycle become rarer. The values are averaged over the ten simulation and as the simulation go through learning. 

With an average total of $628 \pm 283$ reference potential evaluations, the cost of the OTF-ART is between that of ML-MD and OTF-MDART. Along pathways, the average energy error for these potentials is of $0.22\pm0.03$~meV/atom, on par with OTF-MDART potential based on low-temperature ML-MD fitting, and 49~\% lower than the 300~K ML-MD potential. Errors on forces, at 0.011$\pm$0.001~eV/\r{A}, are in between ML-MD (0.012~eV/\r{A}) and OTF-MDART (0.010~eV/\r{A}) at low training temperature. Comparing with the 2700~K potential fitting in MD, OFT-ART error is  57~\% lower than ML-MD (0.026~eV/\r{A}) and 36~\% lower than OTF-MDART (0.017~eV/\r{A}).

Focusing on barrier energy, the average error is $0.039\pm0.008$~eV~(see Fig.~\ref{fig:si-barrier_error-all}), about 2.5~\% lower than OTF-MDART and 30.3~\% better than ML-MD. The error of $0.072\pm0.006$~\r{A} on the converged saddle position is similar to the $0.072\pm0.010$~\r{A} obtained with OTF-MDART and 37~\% lower than with  ML-MD ($0.114$~\r{A}).

\subsection{Reproducing the dominant diffusion mechanism}

\begin{figure}[h]
\centering
\includegraphics[width=\linewidth]{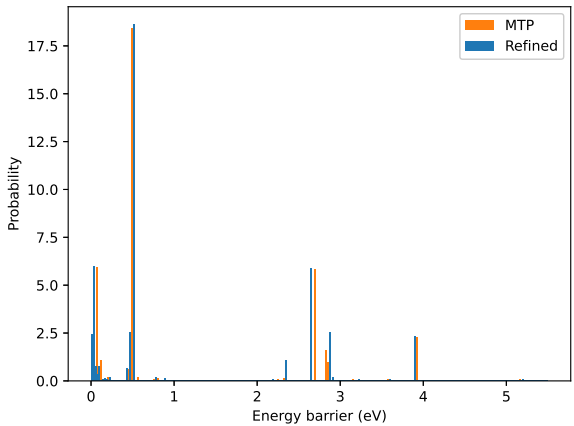}
\caption{\label{fig:histogram_art-si} ARTn-generated energy barrier distributions for vacancy-diffusion events in Si, including direct barrier (from ground state) and inverse barriers (from excited states), as generated with the MTP model (orange) and re-converged using the reference model (blue) from saddles and minima position originally found with the MTP model.}
\end{figure}

The exploration of the energy landscape around the vacancy leads to a generation of wide range of activated mechanisms and associated barriers (both forward, associated with the diffusion of the vacancy, and backward, from the final minima back to the saddle point). Fig.~\ref{fig:histogram_art-si} presents the complete distribution of generated direct and inverse barriers connected to the ground state. The peak near 0~eV (around $10^{-2}$ to $10^{-1}$~eV) is associated with the inverse barrier to to direct saddle at 2.38, 2.70~eV and higher (up to 5.5~eV), except for the inverse 0.45~eV barrier which is linked to the 2.87~eV direct barrier. Direct barriers at 0.51~eV represent symmetric first neighbor vacancy diffusion while barriers at 2.38 and 2.70~eV are associated with more complex vacancy diffusion mechanism~\cite{PhysRevB.70.205202}. Events with barriers at 2.38, 2.70~eV, for example, involve a vacancy diffusion through complex bond-exchanges. Spectator events~\cite{kumeda2001transition} where the diamond network around the vacancy is transformed by a bond switching are also generated. This mechanism was proposed by Wooten, Winer, and Weaire (WWW) to describe the amorphization of silicon~\cite{PhysRevLett.54.1392}. The main spectator event occurs as two neighbors of the vacancy are pushed together allowing the creation of a bound associated with the 2.87~eV barrier. Other mechanisms involve strong lattice distortion and bond formation not involving direct neighbors of the vacancy with very high energy barriers~\cite{PhysRevB.70.205202} of in between 3.2 and 4.0~eV.

\begin{table}[h]
\caption{\label{tab:si-barrier_error-per-0.51}Average energy barrier errors and mean saddle position error on the 0.51~eV vacancy diffusion for Si. The average error for ML-MD and OTF-MDART training is taken over all temperature sets. The standard deviation computed on all temperature sets.}
\begin{ruledtabular}
\begin{tabular}{l c c c r}
\multicolumn{1}{l}{Errors} & \multicolumn{1}{c}{} & ML-MD           & OTF-MDART       & OTF-ART         \\ \hline \\
$\Delta \delta E_{barrier}$ (eV) &                & 0.026$\pm$0.015 & 0.022$\pm$0.011 & 0.019$\pm$0.005 \\
$\Delta X_{\text{conv}}$ (\r{A}) &                & 0.088$\pm$0.036 & 0.040$\pm$0.017 & 0.047$\pm$0.018 \\ 
\end{tabular}
\end{ruledtabular}
\end{table}

Since vacancy diffusion for this system is dominated by a 0.51~eV single barrier mechanism, with the next barrier at 2.35~eV, an accurate description of the dominant mechanism is essential to correctly capture defect kinetics in Si.  Tab.~\ref{tab:si-barrier_error-per-0.51} presents the error on this barrier for the three approaches described above. With an error of 0.019$\pm$0.005~eV, a relative error of 3.7~\%, OTF-ART offers the closest reproduction of the reference barrier, followed by OTF-MDART and ML-MD, with a respective error of 0.022$\pm$0.011 (relative error of 4.3~\%) and 0.026$\pm$0.015 (5.1~\%). Overall, the error on energy barrier is lower than that on the total energy presented above (0.046$\pm$0.006~eV for OTF-ART, for example), due to a partial error cancellation associated with energy difference taken to measure the barrier. 

The validity of the barrier is also measured by the precision on the saddle geometry. For the 0.51~eV barrier, ML-MD converges with an error on the position of 0.088$\pm$0.036\r{A}, with OTF-MDART and OTF-ART giving error almost 50~\% lower, at 0.040$\pm$0.017\r{A} and 0.047$\pm$0.018\r{A} respectively.

\subsection{SiGe system}

\begin{figure}[h]
\centering
\includegraphics[width=\linewidth]{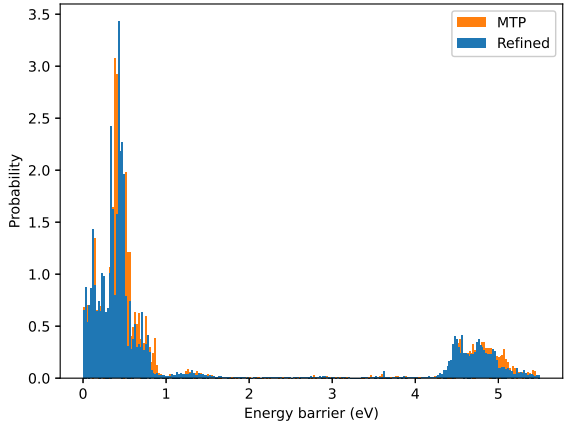}
\caption{\label{fig:histogram_art-sige} SiGe barrier histogram, including direct barrier (from a states accepted) and inverse barriers (from excited states), as found on-the-fly by the MTP model(orange) and re-converge by the reference model(blue) from saddles and minimums position originally given by MTP.}
\end{figure}

\begin{figure}[h]
\centering
\includegraphics[width=\linewidth]{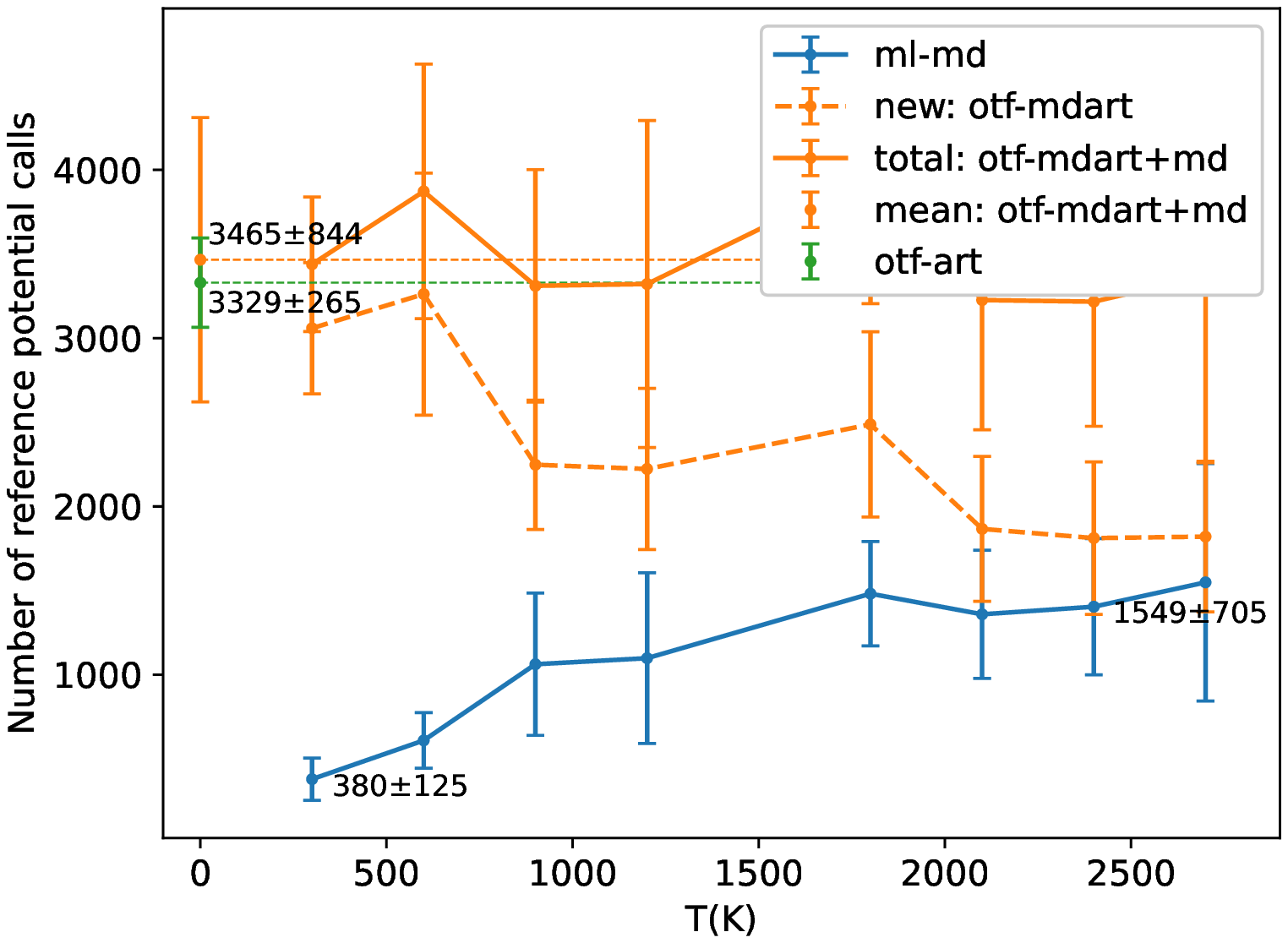}
\caption{\label{fig:sige-counts} Number of calls to the reference potential for each of the OTF machine-learned potentials developed for SiGe as a function of the temperature referring to the one used during MD training. Since configurations are relaxed to zero K in ARTn simulations, there is no associated temperature for this procedure. Vertical bars represent the standard deviation computed on ten independent realisations.}
\end{figure}

\begin{figure}[h]
\centering
\includegraphics[width=\linewidth]{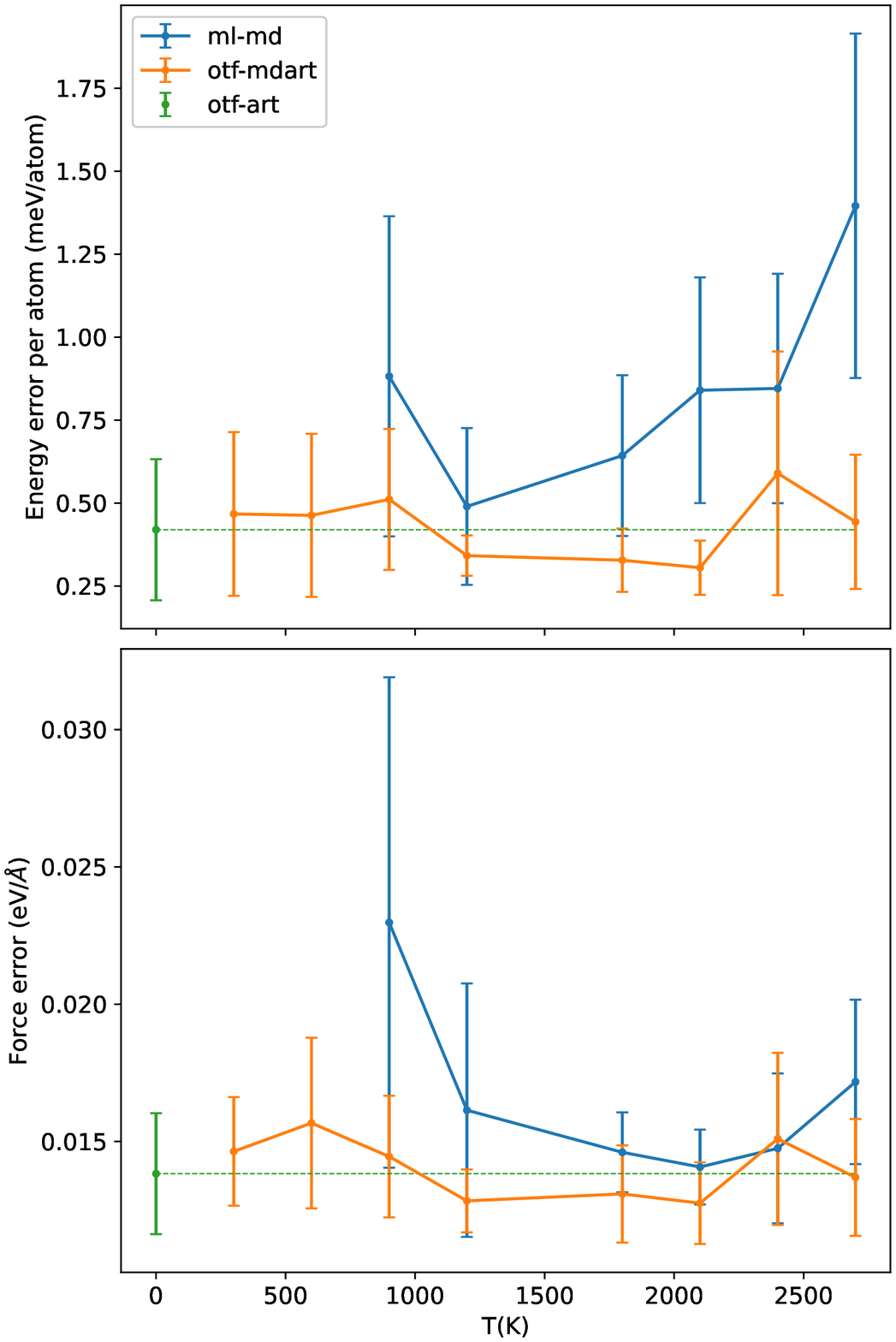}
\caption{\label{fig:sige-error_mean_valid} Average energy (top) and mean absolute forces (bottom) errors for SiGe measured over all configurations generated along pathways in ARTn for the three approaches. Temperature refers to the one used during MD training. Vertical bars represent the standard deviation computed on ten independent realisations.}
\end{figure}

\begin{table}[h]
\caption{\label{tab:sige-barr-pos-error-per}Average energy barrier errors and mean saddle position error on all barriers for SiGe. The average error for ML-MD and OTF-MDART training is taken over all temperature sets. The standard deviation computed on all temperature sets.}
\begin{ruledtabular}
\begin{tabular}{l c c c r}
\multicolumn{1}{c}{Errors} & \multicolumn{1}{c}{} & ML-MD           & OTF-MDART       & OTF-ART         \\ \hline \\
$\Delta \delta E_{barrier}$ (eV) &                & 0.082$\pm$0.024 & 0.072$\pm$0.014 & 0.066$\pm$0.015 \\
$\Delta X_{\text{conv}}$ (\r{A}) &                & 0.091$\pm$0.020 & 0.076$\pm$0.013 & 0.070$\pm$0.014 \\
\end{tabular}
\end{ruledtabular}
\end{table}

Having shown the interest of developing a specific potential by applying on-the-fly learning directly to activated events on a simple system such as c-Si with a vacancy, we test this approach with a more complex alloy with the same overall reference potential to facilitate comparison. Starting from a ordered zincblende structure, the diffusion of a vacancy creates chemical disorder that complexifies the landscape visited as shown by the continuous distribution of activated barriers, including both direct and inverse barriers, found as the vacancy diffuses~(Fig.~\ref{fig:histogram_art-sige}); we note that the lowest barrier for a vacancy diffusing is around 0.6~eV, with lower barriers associated, as for Si, with reverse jumps from metastable states. The energy barrier distribution for a vacancy diffusing in SiGe~(Fig.~\ref{fig:histogram_art-sige}) is much more complex than for Si due to the chemical disorder that builds as the vacancy diffuses. 

As stated in the methodology, the additional complexity of the system imposes a richer machined-learning potential, with a larger set of parameters to encompass the greater diversity in the components and the configurations, due to chemical disorder. Combined, these two levels of complexity (set of parameters and configurational) result in an overall higher numbers of calls to the reference potential as compared to Si, irrespective of the approach used (see Fig.~\ref{fig:sige-counts} (SiGe) vs. Fig.~\ref{fig:si-counts}(Si)): while ML-MD requires between 380 evaluations of the reference potential at 300~K and 1549 at 2700~K, OTF-MDART needs a total of around 3465 calculations of the reference potential, irrespective of the temperature as original ML-MD configurations are progressively removed from the training set. This efforts results in a number of calls to the reference potential for OTF-MDART 4~\% higher than with OTF-ART (3329 on average). To reduce computational costs, we omit the 1500~K run, as statistical behavior is smooth in this temperature region.

\begin{figure}[h]
\centering
\includegraphics[width=\linewidth]{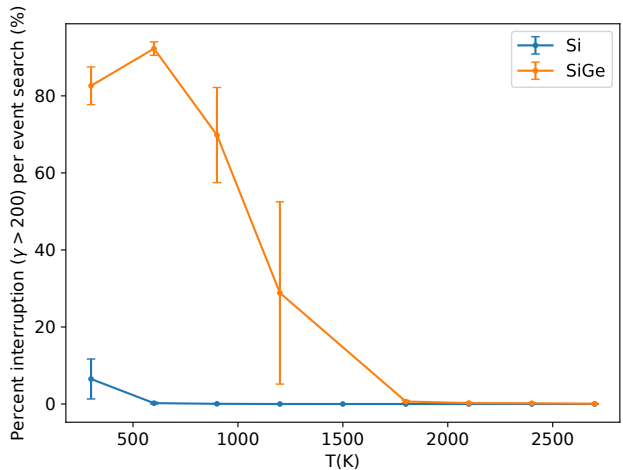}
\caption{\label{fig:failure} Percentage of search interruptions during ML-MD potential evaluation in ARTn ($\gamma > 200$) for Si and SiGe as function of ML-MD training temperature. Vertical bars represent the standard deviation computed on ten independent realisations.}
\end{figure}

To disentangle the two contributions, we compare with the cost of fitting a Si potential with the same level 20 potential as used for SiGe. Following the full OTF-ART procedure, creating a Si MLP requires 2926 calls to the reference potential. The intrinsic complexity of the landscape contributes therefore to about a 14~\% increase of the Si baseline calls count. In terms of accuracy, the  Si MLP level 20 leads to an average error on energy of 0.1~meV/atom, about 50~\% lower than with the level 16 potential described above (0.22~meV/atom). For SiGe, this error is (0.42~meV/atom), two times higher than for Si MLP level 16 and four times that of Si MLP level 20. 

This can be understood by the number of different configurations visited: as opposed to the Si system where each initial minimum is identical (as the vacancy moves in an otherwise perfect elemental crystal), the binary system is transformed as the vacancy diffuses, as the chemical order is slowly destroyed: each of the 24 ARTn parallel trajectories used to define the potential over 1500 events evolves independently according to a probability given by the Metropolis algorithm with a fictitious temperature (since network itself is structurally at 0K) of 0.5~eV (Eq.~\ref{eq:metropolis}), providing a rich range of local environments. 

Fitting a potential is clearly harder: with the parameters used --- when a configuration graded at $\gamma > 200$ is encountered, the ARTn event search is stopped ---, no event could be generated using the ML-MD potential at 300~K and 600~K, which explains the absence of data for this temperatures in Fig.~\ref{fig:sige-error_mean_valid} and~\ref{tab:sige-barr-pos-error-per}.  For SiGe, the error on energy (see Fig.~\ref{fig:sige-error_mean_valid}) with the ML-MD at 900~K and above ranges from 0.5~meV/atom to 1.4~meV/atom, as a function of temperature. On average, these errors are between 14~\% and 69~\% lower with OTF-MDART or OTF-ART at around 0.43~meV/atom.

The  OTF-ART approach gives an error in energy barrier of $0.066\pm0.015$~eV which represent a 19.5~\% and 8.3~\% lower error from the ML-MD ($0.082\pm0.024$~eV) and OTF-MDART ($0.072\pm0.014$~eV) respectively~(Tab.~\ref{tab:sige-barr-pos-error-per}). The errors on the converged saddle position for OTF-ART and OTF-MDART are similar at $0.070\pm0.014$~\r{A} and $0.076\pm0.013$~\r{A}, respectively, and represent a 23~\% lower error than with  ML-MD ($0.092$~\r{A}). This accuracy is similar to that obtained with Si, in contrast to total energy and energy barrier errors. 

We note that the advantage of ML-MD for SiGe is overstated as shown by the proportion of events generated with ML-MD potential that are interrupted due to a too large extrapolation grade, $\gamma > 200$ for both SiGe and Si (Fig.~\ref{fig:failure}): for SiGe between 85~\% and 30~\%  of events are aborted between 300~K and 1200~K, respectively. This proportion falls to zero percent failure at 1800~K.

\section{Discussion}

We compare three approaches aimed at the construction of potentials with machine learning on-the-fly for the exploration of activated mechanism of the potential energy landscape. We evaluate these by computing their efficiency at reproducing  the energy landscape around a vacancy in two systems, a relatively simple Si diamond system (Fig.~\ref{fig:histogram_art-si}) and a more complex SiGe zincblende system that disorders under vacancy diffusion  (Fig.~\ref{fig:histogram_art-sige}), both described with a reference empirical potential to allow for better statistical analysis. 

The first approach, which sets the comparison level, constructs a more general machine learning potential with molecular dynamics (ML-MD), the second on-the-fly adjusts this this generated potential, during the search for activated events using ARTn, while the third approaches constructs a specifically on-the-fly trained a potential during search of activated events (OTF-ART). The efficiency of these three procedures is measured on the quality of the reproduction of the reference potential during the search for activated event.

The baseline, defined by the ML-MD, is competitive with previously published work. Energy errors for the more standard ML MD approach with a level 16 potential range from $0.44\pm0.36$~meV/atom at 300~K to $5.1\pm1.7$~meV/atom at 2700~K (Fig.~\ref{fig:si-error_mean_valid}), an order of magnitude lower or similar than the 4~meV/atom on an MTP potential of level 24 for Si obtained by Zuo \textit{et al.}~\cite{zuo2020performance}, with the difference explained by the fact that activated events involve local deformations from a zero-temperature crystal with a vacancy and that DFT potentials are more difficult to fit than empirical ones~\cite{novikov2020mlip}. \textcolor{black}{That is, using of DFT, we would expect that these conclusions hold as we are looking at difference between our three approaches, not claiming specific accuracy of these.}

Similarly, the relative energy error on the dominant 0.51~eV diffusion barrier for SW Si is of 5.1~\% (0.026~eV) with the ML-MD approach and 3.7~\% (0.019~eV) with the OTF-ART. Using the same MTP potential trained using an OTF MD with an \textit{ab initio} reference potential, Novoselov \textit{et al.} find a 0.20~eV barrier for vacancy diffusion in Si as compared with 0.18~eV with the reference potential, an error of 0.02~eV or a 10.0~\% relative error. 

Overall, the ML-MD approach, especially when run at temperatures between 900 and 1800~K, can generate a generic ML potential with reasonable precision for describing activated mechanisms in Si and SiGe. Developing a more specific OTF potential, generated directly with ARTn on activated trajectories, however, offers a more accurate description of both the energy and geometry at the barriers. 

It is possible to recover this precision by adjusting the original MD potential during ARTn runs, however, this increases the number of calls to the reference potential, raising the total costs beyond that of OTF-ART while largely erasing work made during ML-MD training phase: for Si, between 300 and 1200~K, none of the ML-MD configurations are retained while around 1.3 to  12.5~\% are retained for the potential trained in range of 1500 and 2700~K (Fig.~\ref{fig:origin}, right-hand axis, orange line), but at the cost of lowering the precision on barriers. 

Moving to a more complex system, such an \textcolor{black}{evolving} binary alloy, increases the overall cost of the procedure in terms of calls to the reference potential, as more parameters need to be fit. Here also, the gain on using a specific potential constructed from ARTn trajectories is notable, both in the average errors and their fluctuations. Indeed, the ML-MD potential presents considerable instabilities while generating activated trajectories as it can be see by the number of configurations considered out-side of the potential's scope ($\gamma > 200$), see Fig.~\ref{fig:failure}. \textcolor{black}{This can also be see by comparing the number of evaluations of the reference potential as function of the temperature for OTF-MDART; for Si this  number  decreases while, for SiGe, the total number of evaluations of the reference potential remains constant. This mean that the potential requires significant re-adjustment, even from its high-temperature training, to adapt to the created disorder of the evolving SiGe system.}

\section*{Conclusion}

We compare the advantage of using a more general vs specific machine-learned potential (MLP) to describe activated mechanisms in solid. To do so, we generate first an MLP constructed with the Moment Tensor Potential formalism~\cite{novikov2020mlip,shapeev2016moment} to replicate Stillinger-Weber potential for Si and SiGe crystals with a single vacancy using a standard molecular dynamics procedure (MD-ML). 

Comparing the quality of the reproduction of activated mechanisms with a ML potential further refined during an activation-relaxation technique nouveau sampling of the energy landscape and a potential unique constructed on-the-fly within ARTn, we show that while a general potential can deliver high accuracy for both the barrier geometries and their related energies, error and fluctuations around the average value are significantly lowered by constructing a specific potential, with a number of calls to the reference potential that is lower than a combined approach (MD + ARTn) for a similar precision. 

The advantage of using a specific potential remains when looking at more complex materials, such the SiGe alloys considered here, even though the advantage in terms of calls to the reference is strongly reduced. 
The next steps will involve applying this strategy \textcolor{black}{with DFT as reference potential} to attack problems that have long been out of reach of computational materials sciences, allowing a much closer connection between modeling and experience.

\section{Supplementary Material}
See the supplementary material for a figure of the relation between accuracy and the level of the MTP potential for Si. Also present are figures displaying the accuracy of the saddle position for the 0.51~eV barrier in Si and all barriers in SiGe.

\section{Code and data availability}

The ARTn packages as well as the data reported here are distributed freely. Please contact Normand Mousseau \\ (normand.mousseau@umontreal.ca).

\begin{acknowledgments}
This project is supported through a Discovery grant from the Natural Science and Engineering Research Council of Canada (NSERC). Karl-Étienne Bolduc is grateful to NSERC and IVADO for summer scholarchips. We are grateful to Calcul Québec and Compute Canada for generous allocation of computational resources. 
\end{acknowledgments}

% \nocite{*}
\bibliography{aipsamp}% Produces the bibliography via BibTeX.

\providecommand{\noopsort}[1]{}\providecommand{\singleletter}[1]{#1}%
\begin{thebibliography}{37}%
\makeatletter
\providecommand \@ifxundefined [1]{%
 \@ifx{#1\undefined}
}%
\providecommand \@ifnum [1]{%
 \ifnum #1\expandafter \@firstoftwo
 \else \expandafter \@secondoftwo
 \fi
}%
\providecommand \@ifx [1]{%
 \ifx #1\expandafter \@firstoftwo
 \else \expandafter \@secondoftwo
 \fi
}%
\providecommand \natexlab [1]{#1}%
\providecommand \enquote  [1]{``#1''}%
\providecommand \bibnamefont  [1]{#1}%
\providecommand \bibfnamefont [1]{#1}%
\providecommand \citenamefont [1]{#1}%
\providecommand \href@noop [0]{\@secondoftwo}%
\providecommand \href [0]{\begingroup \@sanitize@url \@href}%
\providecommand \@href[1]{\@@startlink{#1}\@@href}%
\providecommand \@@href[1]{\endgroup#1\@@endlink}%
\providecommand \@sanitize@url [0]{\catcode `\\12\catcode `\$12\catcode
  `\&12\catcode `\#12\catcode `\^12\catcode `\_12\catcode `\%12\relax}%
\providecommand \@@startlink[1]{}%
\providecommand \@@endlink[0]{}%
\providecommand \url  [0]{\begingroup\@sanitize@url \@url }%
\providecommand \@url [1]{\endgroup\@href {#1}{\urlprefix }}%
\providecommand \urlprefix  [0]{URL }%
\providecommand \Eprint [0]{\href }%
\providecommand \doibase [0]{http://dx.doi.org/}%
\providecommand \selectlanguage [0]{\@gobble}%
\providecommand \bibinfo  [0]{\@secondoftwo}%
\providecommand \bibfield  [0]{\@secondoftwo}%
\providecommand \translation [1]{[#1]}%
\providecommand \BibitemOpen [0]{}%
\providecommand \bibitemStop [0]{}%
\providecommand \bibitemNoStop [0]{.\EOS\space}%
\providecommand \EOS [0]{\spacefactor3000\relax}%
\providecommand \BibitemShut  [1]{\csname bibitem#1\endcsname}%
\let\auto@bib@innerbib\@empty
%</preamble>
\bibitem [{\citenamefont {Kohn}\ and\ \citenamefont
  {Sham}(1965)}]{PhysRev.140.A1133}%
  \BibitemOpen
  \bibfield  {author} {\bibinfo {author} {\bibfnamefont {W.}~\bibnamefont
  {Kohn}}\ and\ \bibinfo {author} {\bibfnamefont {L.~J.}\ \bibnamefont
  {Sham}},\ }\bibfield  {title} {\enquote {\bibinfo {title} {Self-consistent
  equations including exchange and correlation effects},}\ }\href {\doibase
  10.1103/PhysRev.140.A1133} {\bibfield  {journal} {\bibinfo  {journal} {Phys.
  Rev.}\ }\textbf {\bibinfo {volume} {140}},\ \bibinfo {pages} {A1133--A1138}
  (\bibinfo {year} {1965})}\BibitemShut {NoStop}%
\bibitem [{\citenamefont {Lindahl}(2008)}]{lindahl2008molecular}%
  \BibitemOpen
  \bibfield  {author} {\bibinfo {author} {\bibfnamefont {E.~R.}\ \bibnamefont
  {Lindahl}},\ }\bibfield  {title} {\enquote {\bibinfo {title} {Molecular
  dynamics simulations},}\ }in\ \href@noop {} {\emph {\bibinfo {booktitle}
  {Molecular modeling of proteins}}}\ (\bibinfo  {publisher} {Springer},\
  \bibinfo {year} {2008})\ pp.\ \bibinfo {pages} {3--23}\BibitemShut {NoStop}%
\bibitem [{\citenamefont {Voter}\ and\ \citenamefont
  {Doll}(1984)}]{voter1984transition}%
  \BibitemOpen
  \bibfield  {author} {\bibinfo {author} {\bibfnamefont {A.~F.}\ \bibnamefont
  {Voter}}\ and\ \bibinfo {author} {\bibfnamefont {J.~D.}\ \bibnamefont
  {Doll}},\ }\bibfield  {title} {\enquote {\bibinfo {title} {Transition state
  theory description of surface self-diffusion: Comparison with classical
  trajectory results},}\ }\href@noop {} {\bibfield  {journal} {\bibinfo
  {journal} {The Journal of chemical physics}\ }\textbf {\bibinfo {volume}
  {80}},\ \bibinfo {pages} {5832--5838} (\bibinfo {year} {1984})}\BibitemShut
  {NoStop}%
\bibitem [{\citenamefont {Voter}(2007)}]{voter2007introduction}%
  \BibitemOpen
  \bibfield  {author} {\bibinfo {author} {\bibfnamefont {A.~F.}\ \bibnamefont
  {Voter}},\ }\bibfield  {title} {\enquote {\bibinfo {title} {Introduction to
  the kinetic monte carlo method},}\ }in\ \href@noop {} {\emph {\bibinfo
  {booktitle} {Radiation effects in solids}}}\ (\bibinfo  {publisher}
  {Springer},\ \bibinfo {year} {2007})\ pp.\ \bibinfo {pages}
  {1--23}\BibitemShut {NoStop}%
\bibitem [{\citenamefont {Henkelman}\ and\ \citenamefont
  {J{\'o}nsson}(2001)}]{henkelman2001}%
  \BibitemOpen
  \bibfield  {author} {\bibinfo {author} {\bibfnamefont {G.}~\bibnamefont
  {Henkelman}}\ and\ \bibinfo {author} {\bibfnamefont {H.}~\bibnamefont
  {J{\'o}nsson}},\ }\bibfield  {title} {\enquote {\bibinfo {title} {Long time
  scale kinetic monte carlo simulations without lattice approximation and
  predefined event table},}\ }\href {\doibase 10.1063/1.1415500} {\bibfield
  {journal} {\bibinfo  {journal} {The Journal of Chemical Physics}\ }\textbf
  {\bibinfo {volume} {115}},\ \bibinfo {pages} {9657--9666} (\bibinfo {year}
  {2001})},\ \Eprint {http://arxiv.org/abs/https://doi.org/10.1063/1.1415500}
  {https://doi.org/10.1063/1.1415500} \BibitemShut {NoStop}%
\bibitem [{\citenamefont {El-Mellouhi}, \citenamefont {Mousseau},\ and\
  \citenamefont {Lewis}(2008)}]{el-mellouhi_kinetic_2008}%
  \BibitemOpen
  \bibfield  {author} {\bibinfo {author} {\bibfnamefont {F.}~\bibnamefont
  {El-Mellouhi}}, \bibinfo {author} {\bibfnamefont {N.}~\bibnamefont
  {Mousseau}}, \ and\ \bibinfo {author} {\bibfnamefont {L.~J.}\ \bibnamefont
  {Lewis}},\ }\bibfield  {title} {\enquote {\bibinfo {title} {Kinetic
  activation-relaxation technique: {An} off-lattice self-learning kinetic
  {Monte} {Carlo} algorithm},}\ }\href {\doibase 10.1103/PhysRevB.78.153202}
  {\bibfield  {journal} {\bibinfo  {journal} {Physical Review B}\ }\textbf
  {\bibinfo {volume} {78}},\ \bibinfo {pages} {153202} (\bibinfo {year}
  {2008})}\BibitemShut {NoStop}%
\bibitem [{\citenamefont {Behler}\ and\ \citenamefont
  {Parrinello}(2007)}]{behler_generalized_2007}%
  \BibitemOpen
  \bibfield  {author} {\bibinfo {author} {\bibfnamefont {J.}~\bibnamefont
  {Behler}}\ and\ \bibinfo {author} {\bibfnamefont {M.}~\bibnamefont
  {Parrinello}},\ }\bibfield  {title} {\enquote {\bibinfo {title} {Generalized
  {Neural}-{Network} {Representation} of {High}-{Dimensional}
  {Potential}-{Energy} {Surfaces}},}\ }\href {\doibase
  10.1103/PhysRevLett.98.146401} {\bibfield  {journal} {\bibinfo  {journal}
  {Phys. Rev. Lett.}\ }\textbf {\bibinfo {volume} {98}},\ \bibinfo {pages}
  {146401} (\bibinfo {year} {2007})}\BibitemShut {NoStop}%
\bibitem [{\citenamefont {Bart{\'o}k}, \citenamefont {Kondor},\ and\
  \citenamefont {Cs{\'a}nyi}(2013)}]{bartok2013representing}%
  \BibitemOpen
  \bibfield  {author} {\bibinfo {author} {\bibfnamefont {A.~P.}\ \bibnamefont
  {Bart{\'o}k}}, \bibinfo {author} {\bibfnamefont {R.}~\bibnamefont {Kondor}},
  \ and\ \bibinfo {author} {\bibfnamefont {G.}~\bibnamefont {Cs{\'a}nyi}},\
  }\bibfield  {title} {\enquote {\bibinfo {title} {On representing chemical
  environments},}\ }\href@noop {} {\bibfield  {journal} {\bibinfo  {journal}
  {Physical Review B}\ }\textbf {\bibinfo {volume} {87}},\ \bibinfo {pages}
  {184115} (\bibinfo {year} {2013})}\BibitemShut {NoStop}%
\bibitem [{\citenamefont {Thompson}\ \emph {et~al.}(2015)\citenamefont
  {Thompson}, \citenamefont {Swiler}, \citenamefont {Trott}, \citenamefont
  {Foiles},\ and\ \citenamefont {Tucker}}]{thompson2015spectral}%
  \BibitemOpen
  \bibfield  {author} {\bibinfo {author} {\bibfnamefont {A.~P.}\ \bibnamefont
  {Thompson}}, \bibinfo {author} {\bibfnamefont {L.~P.}\ \bibnamefont
  {Swiler}}, \bibinfo {author} {\bibfnamefont {C.~R.}\ \bibnamefont {Trott}},
  \bibinfo {author} {\bibfnamefont {S.~M.}\ \bibnamefont {Foiles}}, \ and\
  \bibinfo {author} {\bibfnamefont {G.~J.}\ \bibnamefont {Tucker}},\ }\bibfield
   {title} {\enquote {\bibinfo {title} {Spectral neighbor analysis method for
  automated generation of quantum-accurate interatomic potentials},}\
  }\href@noop {} {\bibfield  {journal} {\bibinfo  {journal} {Journal of
  Computational Physics}\ }\textbf {\bibinfo {volume} {285}},\ \bibinfo {pages}
  {316--330} (\bibinfo {year} {2015})}\BibitemShut {NoStop}%
\bibitem [{\citenamefont {Shapeev}(2016)}]{shapeev2016moment}%
  \BibitemOpen
  \bibfield  {author} {\bibinfo {author} {\bibfnamefont {A.~V.}\ \bibnamefont
  {Shapeev}},\ }\bibfield  {title} {\enquote {\bibinfo {title} {Moment tensor
  potentials: A class of systematically improvable interatomic potentials},}\
  }\href@noop {} {\bibfield  {journal} {\bibinfo  {journal} {Multiscale
  Modeling \& Simulation}\ }\textbf {\bibinfo {volume} {14}},\ \bibinfo {pages}
  {1153--1173} (\bibinfo {year} {2016})}\BibitemShut {NoStop}%
\bibitem [{\citenamefont {Sivaraman}\ \emph {et~al.}(2021)\citenamefont
  {Sivaraman}, \citenamefont {Guo}, \citenamefont {Ward}, \citenamefont {Hoyt},
  \citenamefont {Williamson}, \citenamefont {Foster}, \citenamefont {Benmore},\
  and\ \citenamefont {Jackson}}]{sivaraman2021automated}%
  \BibitemOpen
  \bibfield  {author} {\bibinfo {author} {\bibfnamefont {G.}~\bibnamefont
  {Sivaraman}}, \bibinfo {author} {\bibfnamefont {J.}~\bibnamefont {Guo}},
  \bibinfo {author} {\bibfnamefont {L.}~\bibnamefont {Ward}}, \bibinfo {author}
  {\bibfnamefont {N.}~\bibnamefont {Hoyt}}, \bibinfo {author} {\bibfnamefont
  {M.}~\bibnamefont {Williamson}}, \bibinfo {author} {\bibfnamefont
  {I.}~\bibnamefont {Foster}}, \bibinfo {author} {\bibfnamefont
  {C.}~\bibnamefont {Benmore}}, \ and\ \bibinfo {author} {\bibfnamefont
  {N.}~\bibnamefont {Jackson}},\ }\bibfield  {title} {\enquote {\bibinfo
  {title} {Automated development of molten salt machine learning potentials:
  application to licl},}\ }\href@noop {} {\bibfield  {journal} {\bibinfo
  {journal} {The Journal of Physical Chemistry Letters}\ }\textbf {\bibinfo
  {volume} {12}},\ \bibinfo {pages} {4278--4285} (\bibinfo {year}
  {2021})}\BibitemShut {NoStop}%
\bibitem [{\citenamefont {Kang}, \citenamefont {Shang},\ and\ \citenamefont
  {Liu}(2020)}]{kang2020large}%
  \BibitemOpen
  \bibfield  {author} {\bibinfo {author} {\bibfnamefont {P.-L.}\ \bibnamefont
  {Kang}}, \bibinfo {author} {\bibfnamefont {C.}~\bibnamefont {Shang}}, \ and\
  \bibinfo {author} {\bibfnamefont {Z.-P.}\ \bibnamefont {Liu}},\ }\bibfield
  {title} {\enquote {\bibinfo {title} {Large-scale atomic simulation via
  machine learning potentials constructed by global potential energy surface
  exploration},}\ }\href@noop {} {\bibfield  {journal} {\bibinfo  {journal}
  {Accounts of Chemical Research}\ }\textbf {\bibinfo {volume} {53}},\ \bibinfo
  {pages} {2119--2129} (\bibinfo {year} {2020})}\BibitemShut {NoStop}%
\bibitem [{\citenamefont {Sivaraman}\ \emph {et~al.}(2020)\citenamefont
  {Sivaraman}, \citenamefont {Krishnamoorthy}, \citenamefont {Baur},
  \citenamefont {Holm}, \citenamefont {Stan}, \citenamefont {Cs{\'a}nyi},
  \citenamefont {Benmore},\ and\ \citenamefont
  {V{\'a}zquez-Mayagoitia}}]{sivaraman2020machine}%
  \BibitemOpen
  \bibfield  {author} {\bibinfo {author} {\bibfnamefont {G.}~\bibnamefont
  {Sivaraman}}, \bibinfo {author} {\bibfnamefont {A.~N.}\ \bibnamefont
  {Krishnamoorthy}}, \bibinfo {author} {\bibfnamefont {M.}~\bibnamefont
  {Baur}}, \bibinfo {author} {\bibfnamefont {C.}~\bibnamefont {Holm}}, \bibinfo
  {author} {\bibfnamefont {M.}~\bibnamefont {Stan}}, \bibinfo {author}
  {\bibfnamefont {G.}~\bibnamefont {Cs{\'a}nyi}}, \bibinfo {author}
  {\bibfnamefont {C.}~\bibnamefont {Benmore}}, \ and\ \bibinfo {author}
  {\bibfnamefont {{\'A}.}~\bibnamefont {V{\'a}zquez-Mayagoitia}},\ }\bibfield
  {title} {\enquote {\bibinfo {title} {Machine-learned interatomic potentials
  by active learning: amorphous and liquid hafnium dioxide},}\ }\href@noop {}
  {\bibfield  {journal} {\bibinfo  {journal} {npj Computational Materials}\
  }\textbf {\bibinfo {volume} {6}},\ \bibinfo {pages} {1--8} (\bibinfo {year}
  {2020})}\BibitemShut {NoStop}%
\bibitem [{\citenamefont {Béland}\ \emph {et~al.}(2013)\citenamefont
  {Béland}, \citenamefont {Anahory}, \citenamefont {Smeets}, \citenamefont
  {Guihard}, \citenamefont {Brommer}, \citenamefont {Joly}, \citenamefont
  {Pothier}, \citenamefont {Lewis}, \citenamefont {Mousseau},\ and\
  \citenamefont {Schiettekatte}}]{beland_replenish_2013}%
  \BibitemOpen
  \bibfield  {author} {\bibinfo {author} {\bibfnamefont {L.~K.}\ \bibnamefont
  {Béland}}, \bibinfo {author} {\bibfnamefont {Y.}~\bibnamefont {Anahory}},
  \bibinfo {author} {\bibfnamefont {D.}~\bibnamefont {Smeets}}, \bibinfo
  {author} {\bibfnamefont {M.}~\bibnamefont {Guihard}}, \bibinfo {author}
  {\bibfnamefont {P.}~\bibnamefont {Brommer}}, \bibinfo {author} {\bibfnamefont
  {J.-F.}\ \bibnamefont {Joly}}, \bibinfo {author} {\bibfnamefont {J.-C.}\
  \bibnamefont {Pothier}}, \bibinfo {author} {\bibfnamefont {L.~J.}\
  \bibnamefont {Lewis}}, \bibinfo {author} {\bibfnamefont {N.}~\bibnamefont
  {Mousseau}}, \ and\ \bibinfo {author} {\bibfnamefont {F.}~\bibnamefont
  {Schiettekatte}},\ }\bibfield  {title} {\enquote {\bibinfo {title} {Replenish
  and {Relax}: {Explaining} {Logarithmic} {Annealing} in {Ion}-{Implanted}
  \$c\$-{Si}},}\ }\href {\doibase 10.1103/PhysRevLett.111.105502} {\bibfield
  {journal} {\bibinfo  {journal} {Phys. Rev. Let.}\ }\textbf {\bibinfo {volume}
  {111}},\ \bibinfo {pages} {105502} (\bibinfo {year} {2013})}\BibitemShut
  {NoStop}%
\bibitem [{\citenamefont {Osetsky}\ \emph {et~al.}(2018)\citenamefont
  {Osetsky}, \citenamefont {Beland}, \citenamefont {Barashev},\ and\
  \citenamefont {Zhang}}]{osetsky2018existence}%
  \BibitemOpen
  \bibfield  {author} {\bibinfo {author} {\bibfnamefont {Y.~N.}\ \bibnamefont
  {Osetsky}}, \bibinfo {author} {\bibfnamefont {L.~K.}\ \bibnamefont {Beland}},
  \bibinfo {author} {\bibfnamefont {A.~V.}\ \bibnamefont {Barashev}}, \ and\
  \bibinfo {author} {\bibfnamefont {Y.}~\bibnamefont {Zhang}},\ }\bibfield
  {title} {\enquote {\bibinfo {title} {On the existence and origin of sluggish
  diffusion in chemically disordered concentrated alloys},}\ }\href@noop {}
  {\bibfield  {journal} {\bibinfo  {journal} {Current Opinion in Solid State
  and Materials Science}\ }\textbf {\bibinfo {volume} {22}},\ \bibinfo {pages}
  {65--74} (\bibinfo {year} {2018})}\BibitemShut {NoStop}%
\bibitem [{\citenamefont {Restrepo}\ \emph {et~al.}(2018)\citenamefont
  {Restrepo}, \citenamefont {Mousseau}, \citenamefont {Trochet}, \citenamefont
  {El-Mellouhi}, \citenamefont {Bouhali},\ and\ \citenamefont
  {Becquart}}]{restrepo2018carbon}%
  \BibitemOpen
  \bibfield  {author} {\bibinfo {author} {\bibfnamefont {O.~A.}\ \bibnamefont
  {Restrepo}}, \bibinfo {author} {\bibfnamefont {N.}~\bibnamefont {Mousseau}},
  \bibinfo {author} {\bibfnamefont {M.}~\bibnamefont {Trochet}}, \bibinfo
  {author} {\bibfnamefont {F.}~\bibnamefont {El-Mellouhi}}, \bibinfo {author}
  {\bibfnamefont {O.}~\bibnamefont {Bouhali}}, \ and\ \bibinfo {author}
  {\bibfnamefont {C.~S.}\ \bibnamefont {Becquart}},\ }\bibfield  {title}
  {\enquote {\bibinfo {title} {Carbon diffusion paths and segregation at
  high-angle tilt grain boundaries in $\alpha$-fe studied by using a kinetic
  activation-relation technique},}\ }\href@noop {} {\bibfield  {journal}
  {\bibinfo  {journal} {Physical Review B}\ }\textbf {\bibinfo {volume} {97}},\
  \bibinfo {pages} {054309} (\bibinfo {year} {2018})}\BibitemShut {NoStop}%
\bibitem [{\citenamefont {Barkema}\ and\ \citenamefont
  {Mousseau}(1996)}]{barkema_event-based_1996}%
  \BibitemOpen
  \bibfield  {author} {\bibinfo {author} {\bibfnamefont {G.~T.}\ \bibnamefont
  {Barkema}}\ and\ \bibinfo {author} {\bibfnamefont {N.}~\bibnamefont
  {Mousseau}},\ }\bibfield  {title} {\enquote {\bibinfo {title} {Event-{Based}
  {Relaxation} of {Continuous} {Disordered} {Systems}},}\ }\href {\doibase
  10.1103/PhysRevLett.77.4358} {\bibfield  {journal} {\bibinfo  {journal}
  {Physical Review Letters}\ }\textbf {\bibinfo {volume} {77}},\ \bibinfo
  {pages} {4358--4361} (\bibinfo {year} {1996})}\BibitemShut {NoStop}%
\bibitem [{\citenamefont {Malek}\ and\ \citenamefont
  {Mousseau}(2000{\natexlab{a}})}]{malek_dynamics_2000}%
  \BibitemOpen
  \bibfield  {author} {\bibinfo {author} {\bibfnamefont {R.}~\bibnamefont
  {Malek}}\ and\ \bibinfo {author} {\bibfnamefont {N.}~\bibnamefont
  {Mousseau}},\ }\bibfield  {title} {\enquote {\bibinfo {title} {Dynamics of
  {Lennard}-{Jones} clusters: {A} characterization of the activation-relaxation
  technique},}\ }\href {\doibase 10.1103/PhysRevE.62.7723} {\bibfield
  {journal} {\bibinfo  {journal} {Physical Review E}\ }\textbf {\bibinfo
  {volume} {62}},\ \bibinfo {pages} {7723--7728} (\bibinfo {year}
  {2000}{\natexlab{a}})}\BibitemShut {NoStop}%
\bibitem [{\citenamefont {Jay}\ \emph {et~al.}(2022{\natexlab{a}})\citenamefont
  {Jay}, \citenamefont {Gunde}, \citenamefont {Salles}, \citenamefont
  {Poberžnik}, \citenamefont {Martin-Samos}, \citenamefont {Richard},
  \citenamefont {Gironcoli}, \citenamefont {Mousseau},\ and\ \citenamefont
  {Hémeryck}}]{jay_activationrelaxation_2022}%
  \BibitemOpen
  \bibfield  {author} {\bibinfo {author} {\bibfnamefont {A.}~\bibnamefont
  {Jay}}, \bibinfo {author} {\bibfnamefont {M.}~\bibnamefont {Gunde}}, \bibinfo
  {author} {\bibfnamefont {N.}~\bibnamefont {Salles}}, \bibinfo {author}
  {\bibfnamefont {M.}~\bibnamefont {Poberžnik}}, \bibinfo {author}
  {\bibfnamefont {L.}~\bibnamefont {Martin-Samos}}, \bibinfo {author}
  {\bibfnamefont {N.}~\bibnamefont {Richard}}, \bibinfo {author} {\bibfnamefont
  {S.~d.}\ \bibnamefont {Gironcoli}}, \bibinfo {author} {\bibfnamefont
  {N.}~\bibnamefont {Mousseau}}, \ and\ \bibinfo {author} {\bibfnamefont
  {A.}~\bibnamefont {Hémeryck}},\ }\bibfield  {title} {\enquote {\bibinfo
  {title} {Activation–{Relaxation} {Technique}: {An} efficient way to find
  minima and saddle points of potential energy surfaces},}\ }\href {\doibase
  10.1016/j.commatsci.2022.111363} {\bibfield  {journal} {\bibinfo  {journal}
  {Computational Materials Science}\ }\textbf {\bibinfo {volume} {209}},\
  \bibinfo {pages} {111363} (\bibinfo {year} {2022}{\natexlab{a}})}\BibitemShut
  {NoStop}%
\bibitem [{\citenamefont {Truhlar}, \citenamefont {Garrett},\ and\
  \citenamefont {Klippenstein}(1996)}]{truhlar1996current}%
  \BibitemOpen
  \bibfield  {author} {\bibinfo {author} {\bibfnamefont {D.~G.}\ \bibnamefont
  {Truhlar}}, \bibinfo {author} {\bibfnamefont {B.~C.}\ \bibnamefont
  {Garrett}}, \ and\ \bibinfo {author} {\bibfnamefont {S.~J.}\ \bibnamefont
  {Klippenstein}},\ }\bibfield  {title} {\enquote {\bibinfo {title} {Current
  status of transition-state theory},}\ }\href@noop {} {\bibfield  {journal}
  {\bibinfo  {journal} {The Journal of physical chemistry}\ }\textbf {\bibinfo
  {volume} {100}},\ \bibinfo {pages} {12771--12800} (\bibinfo {year}
  {1996})}\BibitemShut {NoStop}%
\bibitem [{\citenamefont {Novikov}\ \emph {et~al.}(2020)\citenamefont
  {Novikov}, \citenamefont {Gubaev}, \citenamefont {Podryabinkin},\ and\
  \citenamefont {Shapeev}}]{novikov2020mlip}%
  \BibitemOpen
  \bibfield  {author} {\bibinfo {author} {\bibfnamefont {I.~S.}\ \bibnamefont
  {Novikov}}, \bibinfo {author} {\bibfnamefont {K.}~\bibnamefont {Gubaev}},
  \bibinfo {author} {\bibfnamefont {E.~V.}\ \bibnamefont {Podryabinkin}}, \
  and\ \bibinfo {author} {\bibfnamefont {A.~V.}\ \bibnamefont {Shapeev}},\
  }\bibfield  {title} {\enquote {\bibinfo {title} {The mlip package: moment
  tensor potentials with mpi and active learning},}\ }\href@noop {} {\bibfield
  {journal} {\bibinfo  {journal} {Machine Learning: Science and Technology}\
  }\textbf {\bibinfo {volume} {2}},\ \bibinfo {pages} {025002} (\bibinfo {year}
  {2020})}\BibitemShut {NoStop}%
\bibitem [{\citenamefont {Novoselov}\ \emph {et~al.}(2019)\citenamefont
  {Novoselov}, \citenamefont {Yanilkin}, \citenamefont {Shapeev},\ and\
  \citenamefont {Podryabinkin}}]{novoselov2019moment}%
  \BibitemOpen
  \bibfield  {author} {\bibinfo {author} {\bibfnamefont {I.}~\bibnamefont
  {Novoselov}}, \bibinfo {author} {\bibfnamefont {A.}~\bibnamefont {Yanilkin}},
  \bibinfo {author} {\bibfnamefont {A.}~\bibnamefont {Shapeev}}, \ and\
  \bibinfo {author} {\bibfnamefont {E.}~\bibnamefont {Podryabinkin}},\
  }\bibfield  {title} {\enquote {\bibinfo {title} {Moment tensor potentials as
  a promising tool to study diffusion processes},}\ }\href@noop {} {\bibfield
  {journal} {\bibinfo  {journal} {Computational Materials Science}\ }\textbf
  {\bibinfo {volume} {164}},\ \bibinfo {pages} {46--56} (\bibinfo {year}
  {2019})}\BibitemShut {NoStop}%
\bibitem [{\citenamefont {Stillinger}\ and\ \citenamefont
  {Weber}(1985)}]{stillinger1985computer}%
  \BibitemOpen
  \bibfield  {author} {\bibinfo {author} {\bibfnamefont {F.~H.}\ \bibnamefont
  {Stillinger}}\ and\ \bibinfo {author} {\bibfnamefont {T.~A.}\ \bibnamefont
  {Weber}},\ }\bibfield  {title} {\enquote {\bibinfo {title} {Computer
  simulation of local order in condensed phases of silicon},}\ }\href@noop {}
  {\bibfield  {journal} {\bibinfo  {journal} {Physical review B}\ }\textbf
  {\bibinfo {volume} {31}},\ \bibinfo {pages} {5262} (\bibinfo {year}
  {1985})}\BibitemShut {NoStop}%
\bibitem [{\citenamefont {Ethier}\ and\ \citenamefont
  {Lewis}(1992)}]{ethier1992epitaxial}%
  \BibitemOpen
  \bibfield  {author} {\bibinfo {author} {\bibfnamefont {S.}~\bibnamefont
  {Ethier}}\ and\ \bibinfo {author} {\bibfnamefont {L.~J.}\ \bibnamefont
  {Lewis}},\ }\bibfield  {title} {\enquote {\bibinfo {title} {Epitaxial growth
  of si1- xgex on si (100) 2$\times$ 1: A molecular-dynamics study},}\
  }\href@noop {} {\bibfield  {journal} {\bibinfo  {journal} {Journal of
  materials research}\ }\textbf {\bibinfo {volume} {7}},\ \bibinfo {pages}
  {2817--2827} (\bibinfo {year} {1992})}\BibitemShut {NoStop}%
\bibitem [{\citenamefont {Zuo}\ \emph {et~al.}(2020)\citenamefont {Zuo},
  \citenamefont {Chen}, \citenamefont {Li}, \citenamefont {Deng}, \citenamefont
  {Chen}, \citenamefont {Behler}, \citenamefont {Cs{\'a}nyi}, \citenamefont
  {Shapeev}, \citenamefont {Thompson}, \citenamefont {Wood} \emph
  {et~al.}}]{zuo2020performance}%
  \BibitemOpen
  \bibfield  {author} {\bibinfo {author} {\bibfnamefont {Y.}~\bibnamefont
  {Zuo}}, \bibinfo {author} {\bibfnamefont {C.}~\bibnamefont {Chen}}, \bibinfo
  {author} {\bibfnamefont {X.}~\bibnamefont {Li}}, \bibinfo {author}
  {\bibfnamefont {Z.}~\bibnamefont {Deng}}, \bibinfo {author} {\bibfnamefont
  {Y.}~\bibnamefont {Chen}}, \bibinfo {author} {\bibfnamefont {J.}~\bibnamefont
  {Behler}}, \bibinfo {author} {\bibfnamefont {G.}~\bibnamefont {Cs{\'a}nyi}},
  \bibinfo {author} {\bibfnamefont {A.~V.}\ \bibnamefont {Shapeev}}, \bibinfo
  {author} {\bibfnamefont {A.~P.}\ \bibnamefont {Thompson}}, \bibinfo {author}
  {\bibfnamefont {M.~A.}\ \bibnamefont {Wood}},  \emph {et~al.},\ }\bibfield
  {title} {\enquote {\bibinfo {title} {Performance and cost assessment of
  machine learning interatomic potentials},}\ }\href@noop {} {\bibfield
  {journal} {\bibinfo  {journal} {The Journal of Physical Chemistry A}\
  }\textbf {\bibinfo {volume} {124}},\ \bibinfo {pages} {731--745} (\bibinfo
  {year} {2020})}\BibitemShut {NoStop}%
\bibitem [{\citenamefont {Podryabinkin}\ and\ \citenamefont
  {Shapeev}(2017)}]{podryabinkin2017active}%
  \BibitemOpen
  \bibfield  {author} {\bibinfo {author} {\bibfnamefont {E.~V.}\ \bibnamefont
  {Podryabinkin}}\ and\ \bibinfo {author} {\bibfnamefont {A.~V.}\ \bibnamefont
  {Shapeev}},\ }\bibfield  {title} {\enquote {\bibinfo {title} {Active learning
  of linearly parametrized interatomic potentials},}\ }\href@noop {} {\bibfield
   {journal} {\bibinfo  {journal} {Computational Materials Science}\ }\textbf
  {\bibinfo {volume} {140}},\ \bibinfo {pages} {171--180} (\bibinfo {year}
  {2017})}\BibitemShut {NoStop}%
\bibitem [{\citenamefont {Podryabinkin}\ \emph {et~al.}(2019)\citenamefont
  {Podryabinkin}, \citenamefont {Tikhonov}, \citenamefont {Shapeev},\ and\
  \citenamefont {Oganov}}]{podryabinkin2019accelerating}%
  \BibitemOpen
  \bibfield  {author} {\bibinfo {author} {\bibfnamefont {E.~V.}\ \bibnamefont
  {Podryabinkin}}, \bibinfo {author} {\bibfnamefont {E.~V.}\ \bibnamefont
  {Tikhonov}}, \bibinfo {author} {\bibfnamefont {A.~V.}\ \bibnamefont
  {Shapeev}}, \ and\ \bibinfo {author} {\bibfnamefont {A.~R.}\ \bibnamefont
  {Oganov}},\ }\bibfield  {title} {\enquote {\bibinfo {title} {Accelerating
  crystal structure prediction by machine-learning interatomic potentials with
  active learning},}\ }\href@noop {} {\bibfield  {journal} {\bibinfo  {journal}
  {Physical Review B}\ }\textbf {\bibinfo {volume} {99}},\ \bibinfo {pages}
  {064114} (\bibinfo {year} {2019})}\BibitemShut {NoStop}%
\bibitem [{\citenamefont {Gubaev}\ \emph {et~al.}(2019)\citenamefont {Gubaev},
  \citenamefont {Podryabinkin}, \citenamefont {Hart},\ and\ \citenamefont
  {Shapeev}}]{gubaev2019accelerating}%
  \BibitemOpen
  \bibfield  {author} {\bibinfo {author} {\bibfnamefont {K.}~\bibnamefont
  {Gubaev}}, \bibinfo {author} {\bibfnamefont {E.~V.}\ \bibnamefont
  {Podryabinkin}}, \bibinfo {author} {\bibfnamefont {G.~L.}\ \bibnamefont
  {Hart}}, \ and\ \bibinfo {author} {\bibfnamefont {A.~V.}\ \bibnamefont
  {Shapeev}},\ }\bibfield  {title} {\enquote {\bibinfo {title} {Accelerating
  high-throughput searches for new alloys with active learning of interatomic
  potentials},}\ }\href@noop {} {\bibfield  {journal} {\bibinfo  {journal}
  {Computational Materials Science}\ }\textbf {\bibinfo {volume} {156}},\
  \bibinfo {pages} {148--156} (\bibinfo {year} {2019})}\BibitemShut {NoStop}%
\bibitem [{\citenamefont {Thompson}\ \emph {et~al.}(2022)\citenamefont
  {Thompson}, \citenamefont {Aktulga}, \citenamefont {Berger}, \citenamefont
  {Bolintineanu}, \citenamefont {Brown}, \citenamefont {Crozier}, \citenamefont
  {in~'t Veld}, \citenamefont {Kohlmeyer}, \citenamefont {Moore}, \citenamefont
  {Nguyen}, \citenamefont {Shan}, \citenamefont {Stevens}, \citenamefont
  {Tranchida}, \citenamefont {Trott},\ and\ \citenamefont {Plimpton}}]{LAMMPS}%
  \BibitemOpen
  \bibfield  {author} {\bibinfo {author} {\bibfnamefont {A.~P.}\ \bibnamefont
  {Thompson}}, \bibinfo {author} {\bibfnamefont {H.~M.}\ \bibnamefont
  {Aktulga}}, \bibinfo {author} {\bibfnamefont {R.}~\bibnamefont {Berger}},
  \bibinfo {author} {\bibfnamefont {D.~S.}\ \bibnamefont {Bolintineanu}},
  \bibinfo {author} {\bibfnamefont {W.~M.}\ \bibnamefont {Brown}}, \bibinfo
  {author} {\bibfnamefont {P.~S.}\ \bibnamefont {Crozier}}, \bibinfo {author}
  {\bibfnamefont {P.~J.}\ \bibnamefont {in~'t Veld}}, \bibinfo {author}
  {\bibfnamefont {A.}~\bibnamefont {Kohlmeyer}}, \bibinfo {author}
  {\bibfnamefont {S.~G.}\ \bibnamefont {Moore}}, \bibinfo {author}
  {\bibfnamefont {T.~D.}\ \bibnamefont {Nguyen}}, \bibinfo {author}
  {\bibfnamefont {R.}~\bibnamefont {Shan}}, \bibinfo {author} {\bibfnamefont
  {M.~J.}\ \bibnamefont {Stevens}}, \bibinfo {author} {\bibfnamefont
  {J.}~\bibnamefont {Tranchida}}, \bibinfo {author} {\bibfnamefont
  {C.}~\bibnamefont {Trott}}, \ and\ \bibinfo {author} {\bibfnamefont {S.~J.}\
  \bibnamefont {Plimpton}},\ }\bibfield  {title} {\enquote {\bibinfo {title}
  {{LAMMPS} - a flexible simulation tool for particle-based materials modeling
  at the atomic, meso, and continuum scales},}\ }\href {\doibase
  10.1016/j.cpc.2021.108171} {\bibfield  {journal} {\bibinfo  {journal} {Comp.
  Phys. Comm.}\ }\textbf {\bibinfo {volume} {271}},\ \bibinfo {pages} {108171}
  (\bibinfo {year} {2022})}\BibitemShut {NoStop}%
\bibitem [{\citenamefont {Jay}\ \emph {et~al.}(2022{\natexlab{b}})\citenamefont
  {Jay}, \citenamefont {Gunde}, \citenamefont {Salles}, \citenamefont
  {Pober{\v{z}}nik}, \citenamefont {Martin-Samos}, \citenamefont {Richard},
  \citenamefont {de~Gironcoli}, \citenamefont {Mousseau},\ and\ \citenamefont
  {H{\'e}meryck}}]{jay2022activation}%
  \BibitemOpen
  \bibfield  {author} {\bibinfo {author} {\bibfnamefont {A.}~\bibnamefont
  {Jay}}, \bibinfo {author} {\bibfnamefont {M.}~\bibnamefont {Gunde}}, \bibinfo
  {author} {\bibfnamefont {N.}~\bibnamefont {Salles}}, \bibinfo {author}
  {\bibfnamefont {M.}~\bibnamefont {Pober{\v{z}}nik}}, \bibinfo {author}
  {\bibfnamefont {L.}~\bibnamefont {Martin-Samos}}, \bibinfo {author}
  {\bibfnamefont {N.}~\bibnamefont {Richard}}, \bibinfo {author} {\bibfnamefont
  {S.}~\bibnamefont {de~Gironcoli}}, \bibinfo {author} {\bibfnamefont
  {N.}~\bibnamefont {Mousseau}}, \ and\ \bibinfo {author} {\bibfnamefont
  {A.}~\bibnamefont {H{\'e}meryck}},\ }\bibfield  {title} {\enquote {\bibinfo
  {title} {Activation--relaxation technique: An efficient way to find minima
  and saddle points of potential energy surfaces},}\ }\href@noop {} {\bibfield
  {journal} {\bibinfo  {journal} {Computational Materials Science}\ }\textbf
  {\bibinfo {volume} {209}},\ \bibinfo {pages} {111363} (\bibinfo {year}
  {2022}{\natexlab{b}})}\BibitemShut {NoStop}%
\bibitem [{\citenamefont {Lanczos}(1950)}]{lanczos1950iteration}%
  \BibitemOpen
  \bibfield  {author} {\bibinfo {author} {\bibfnamefont {C.}~\bibnamefont
  {Lanczos}},\ }\bibfield  {title} {\enquote {\bibinfo {title} {An iteration
  method for the solution of the eigenvalue problem of linear differential and
  integral operators},}\ }\href@noop {} {\  (\bibinfo {year}
  {1950})}\BibitemShut {NoStop}%
\bibitem [{\citenamefont {Malek}\ and\ \citenamefont
  {Mousseau}(2000{\natexlab{b}})}]{malek2000dynamics}%
  \BibitemOpen
  \bibfield  {author} {\bibinfo {author} {\bibfnamefont {R.}~\bibnamefont
  {Malek}}\ and\ \bibinfo {author} {\bibfnamefont {N.}~\bibnamefont
  {Mousseau}},\ }\bibfield  {title} {\enquote {\bibinfo {title} {Dynamics of
  lennard-jones clusters: A characterization of the activation-relaxation
  technique},}\ }\href@noop {} {\bibfield  {journal} {\bibinfo  {journal}
  {Physical Review E}\ }\textbf {\bibinfo {volume} {62}},\ \bibinfo {pages}
  {7723} (\bibinfo {year} {2000}{\natexlab{b}})}\BibitemShut {NoStop}%
\bibitem [{\citenamefont {Bitzek}\ \emph {et~al.}(2006)\citenamefont {Bitzek},
  \citenamefont {Koskinen}, \citenamefont {G{\"a}hler}, \citenamefont
  {Moseler},\ and\ \citenamefont {Gumbsch}}]{bitzek2006structural}%
  \BibitemOpen
  \bibfield  {author} {\bibinfo {author} {\bibfnamefont {E.}~\bibnamefont
  {Bitzek}}, \bibinfo {author} {\bibfnamefont {P.}~\bibnamefont {Koskinen}},
  \bibinfo {author} {\bibfnamefont {F.}~\bibnamefont {G{\"a}hler}}, \bibinfo
  {author} {\bibfnamefont {M.}~\bibnamefont {Moseler}}, \ and\ \bibinfo
  {author} {\bibfnamefont {P.}~\bibnamefont {Gumbsch}},\ }\bibfield  {title}
  {\enquote {\bibinfo {title} {Structural relaxation made simple},}\
  }\href@noop {} {\bibfield  {journal} {\bibinfo  {journal} {Physical review
  letters}\ }\textbf {\bibinfo {volume} {97}},\ \bibinfo {pages} {170201}
  (\bibinfo {year} {2006})}\BibitemShut {NoStop}%
\bibitem [{\citenamefont {Mousseau}\ and\ \citenamefont
  {Barkema}(1999)}]{mousseau_exploring_1999}%
  \BibitemOpen
  \bibfield  {author} {\bibinfo {author} {\bibfnamefont {N.}~\bibnamefont
  {Mousseau}}\ and\ \bibinfo {author} {\bibfnamefont {G.~T.}\ \bibnamefont
  {Barkema}},\ }\bibfield  {title} {\enquote {\bibinfo {title} {Exploring
  {High}‐{Dimensional} {Energy} {Landscapes}},}\ }\href {\doibase
  10.1109/5992.753050} {\bibfield  {journal} {\bibinfo  {journal} {Computing in
  Science \& Engineering}\ }\textbf {\bibinfo {volume} {1}},\ \bibinfo {pages}
  {74--82} (\bibinfo {year} {1999})}\BibitemShut {NoStop}%
\bibitem [{\citenamefont {El-Mellouhi}, \citenamefont {Mousseau},\ and\
  \citenamefont {Ordej\'on}(2004)}]{PhysRevB.70.205202}%
  \BibitemOpen
  \bibfield  {author} {\bibinfo {author} {\bibfnamefont {F.}~\bibnamefont
  {El-Mellouhi}}, \bibinfo {author} {\bibfnamefont {N.}~\bibnamefont
  {Mousseau}}, \ and\ \bibinfo {author} {\bibfnamefont {P.}~\bibnamefont
  {Ordej\'on}},\ }\bibfield  {title} {\enquote {\bibinfo {title} {Sampling the
  diffusion paths of a neutral vacancy in silicon with quantum mechanical
  calculations},}\ }\href {\doibase 10.1103/PhysRevB.70.205202} {\bibfield
  {journal} {\bibinfo  {journal} {Phys. Rev. B}\ }\textbf {\bibinfo {volume}
  {70}},\ \bibinfo {pages} {205202} (\bibinfo {year} {2004})}\BibitemShut
  {NoStop}%
\bibitem [{\citenamefont {Kumeda}, \citenamefont {Wales},\ and\ \citenamefont
  {Munro}(2001)}]{kumeda2001transition}%
  \BibitemOpen
  \bibfield  {author} {\bibinfo {author} {\bibfnamefont {Y.}~\bibnamefont
  {Kumeda}}, \bibinfo {author} {\bibfnamefont {D.~J.}\ \bibnamefont {Wales}}, \
  and\ \bibinfo {author} {\bibfnamefont {L.~J.}\ \bibnamefont {Munro}},\
  }\bibfield  {title} {\enquote {\bibinfo {title} {Transition states and
  rearrangement mechanisms from hybrid eigenvector-following and density
  functional theory.: application to c10h10 and defect migration in crystalline
  silicon},}\ }\href@noop {} {\bibfield  {journal} {\bibinfo  {journal}
  {Chemical physics letters}\ }\textbf {\bibinfo {volume} {341}},\ \bibinfo
  {pages} {185--194} (\bibinfo {year} {2001})}\BibitemShut {NoStop}%
\bibitem [{\citenamefont {Wooten}, \citenamefont {Winer},\ and\ \citenamefont
  {Weaire}(1985)}]{PhysRevLett.54.1392}%
  \BibitemOpen
  \bibfield  {author} {\bibinfo {author} {\bibfnamefont {F.}~\bibnamefont
  {Wooten}}, \bibinfo {author} {\bibfnamefont {K.}~\bibnamefont {Winer}}, \
  and\ \bibinfo {author} {\bibfnamefont {D.}~\bibnamefont {Weaire}},\
  }\bibfield  {title} {\enquote {\bibinfo {title} {Computer generation of
  structural models of amorphous si and ge},}\ }\href {\doibase
  10.1103/PhysRevLett.54.1392} {\bibfield  {journal} {\bibinfo  {journal}
  {Phys. Rev. Lett.}\ }\textbf {\bibinfo {volume} {54}},\ \bibinfo {pages}
  {1392--1395} (\bibinfo {year} {1985})}\BibitemShut {NoStop}%
\end{thebibliography}%

\end{document}

%\preprint{ APL19-AR-11079  }

\title{Evaluating approaches for on-the-fly machine learning interatomic potential for activated mechanisms sampling with the activation-relaxation technique nouveau}
% Force line breaks with \\

\author{Eugène Sanscartier}
\email{eugene.sanscartier@umontreal.ca}
\affiliation{Département de physique and Regroupement québécois sur les matériaux de pointe, Université de Montréal, Case Postale 6128, Succursale Centre-ville, Montréal, Québec H3C 3J7, Canada.}
 
\author{Félix Saint-Denis}
\affiliation{Département de physique and Regroupement québécois sur les matériaux de pointe, Université de Montréal, Case Postale 6128, Succursale Centre-ville, Montréal, Québec H3C 3J7, Canada.}

\author{Karl-Étienne Bolduc}
\affiliation{Département de physique and Regroupement québécois sur les matériaux de pointe, Université de Montréal, Case Postale 6128, Succursale Centre-ville, Montréal, Québec H3C 3J7, Canada.}

\author{Normand Mousseau}
\email{normand.mousseau@umontreal.ca}
\homepage{https://normandmousseau.com}
\affiliation{Département de physique and Regroupement québécois sur les matériaux de pointe, Université de Montréal, Case Postale 6128, Succursale Centre-ville, Montréal, Québec H3C 3J7, Canada.}

\date{\today}% It is always \today, today,

\maketitle
\beginsupplement

\begin{figure}[h]
\centering
\includegraphics[width=\linewidth]{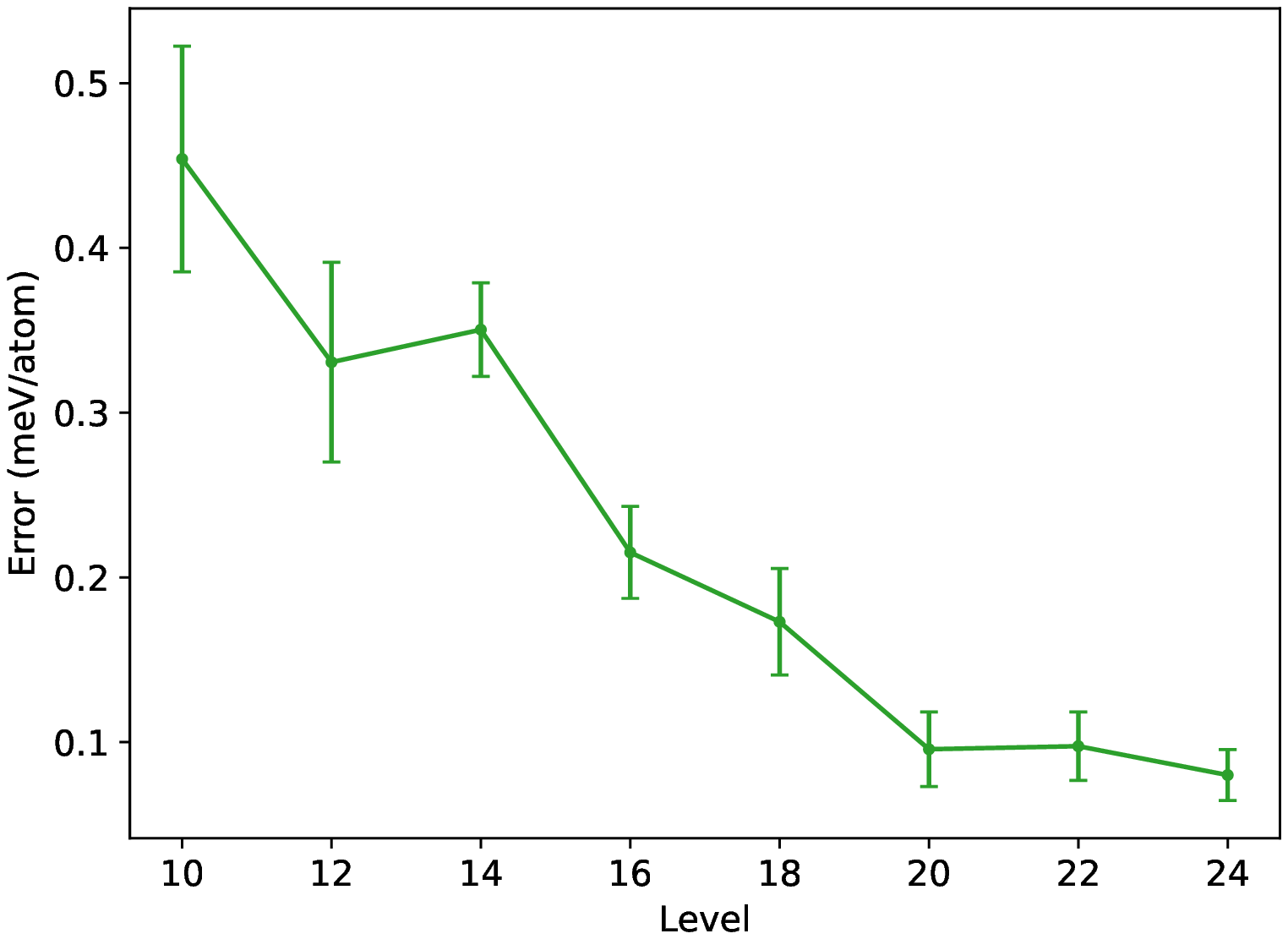}
\caption{\label{fig:si-level-it_error_mean_validation} Energy error per atom averaged  of all configurations generated along pathways in ARTn for Si with the standard OTF-ART approach as function of the number parameter (level). Vertical bars represent the standard deviation computed on ten independent realisations. }
\end{figure}

\begin{figure}[h]
\centering
\includegraphics[width=\linewidth]{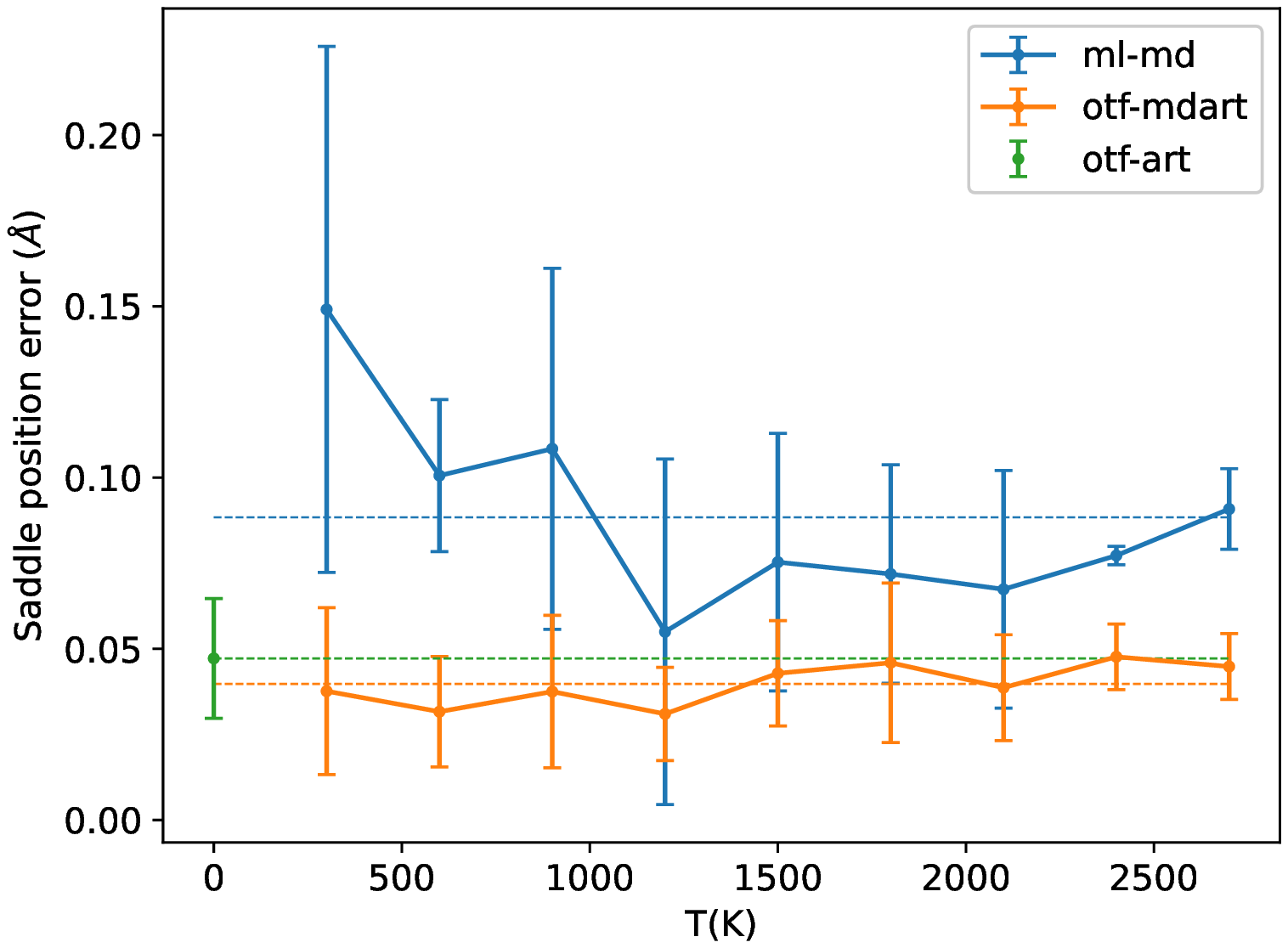}
\caption{\label{fig:saddle-0.51-si} Mean position error on the 0.51~eV vacancy diffusion saddle point for Si. Temperature refers to the one used during MD training. Vertical bars represent the standard deviation computed on ten independent realisations. }
\end{figure}

\begin{figure}[h]
\centering
\includegraphics[width=\linewidth]{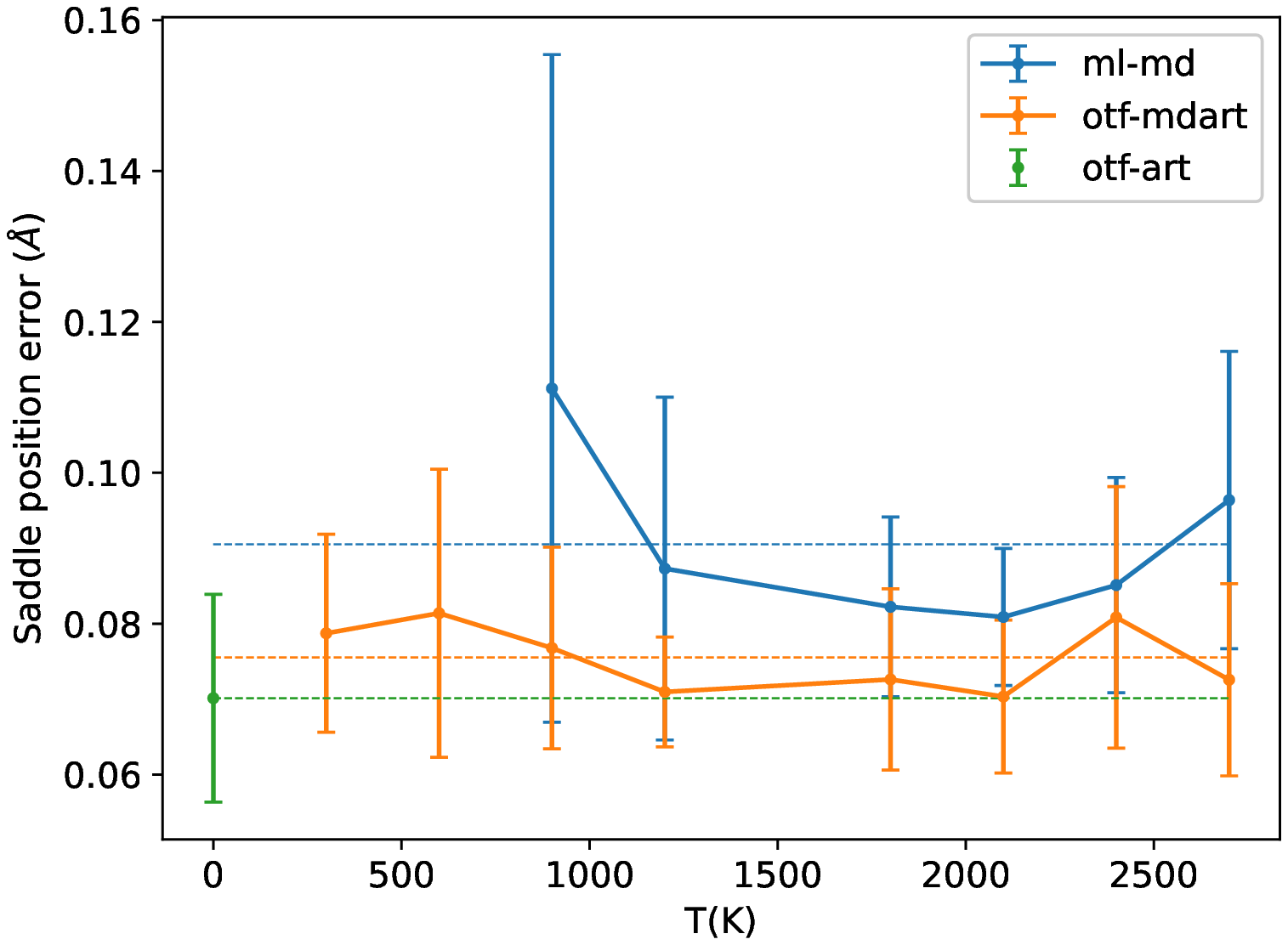}
\caption{\label{fig:saddle-sige} Mean position error on all saddle point for SiGe. Temperature refers to the one used during MD training. Vertical bars represent the standard deviation computed on ten independent realisations. }
\end{figure}